\documentclass[11pt]{article}
\usepackage{epsfig}
\usepackage{axodraw}
\def\simlt{\stackrel{<}{{}_\sim}}
\def\simgt{\stackrel{>}{{}_\sim}}
\usepackage[polish,english,german,french]{babel}
\begin{document}
\selectlanguage{english}

\begin{titlepage}
%\title

\begin{flushright}
IFT-2005/3\\
CERN-PH-TH/2005-026\\
%{\bf \today}
\end{flushright}
\vskip1.0cm

\begin{center}
PHENOMENOLOGICAL GUIDE TO PHYSICS BEYOND THE STANDARD MODEL
\footnote{Lectures given at the Cargese School on String Theory, 
Cargese, 4-16 June 2004}\\
\end{center}
%\author
\vspace{0.5cm}
\begin{center}
Stefan Pokorski\\
Institute for Theoretical Physics, University of Warsaw\\
and\\
Theory Division, CERN, Geneve
\end{center}
\vskip0.5cm

\abstract {Various aspects of physics beyond the Standard Model are discussed 
from the perspective of the fantastic phenomenological success of the Standard 
Model, its simplicity and predictive power.}

%\date{}
%\maketitle
\vskip1.5cm
Content
\begin{enumerate}
\item [1.]Introduction
\item [2.]The Standard Model\\
~~~~~~~~~~2.1 Basic structure (chirality)\\
~~~~~~~~~~2.2 Fermion masses\\
~~~~~~~~~~2.3 Approximate custodial symmetry of the Standard Model\\
~~~~~~~~~~~~~~~~~$\phantom{aaa}$ and the precision electroweak data\\
~~~~~~~~~~2.4 GIM mechanism and the suppression of FCNC and CP\\
~~~~~~~~~~~~~~~~~$\phantom{aaa}$ violating transitions\\
          2.5 Baryon and lepton number conservation
\item [3.]Hints from the Standard Model for its extensions\\
~~~~~~~~~~3.1 Is the effective low energy electroweak theory\\
~~~~~~~~~~~~~~~~~$\phantom{aaa}$ indeed the renormalizable Standard Model?\\
~~~~~~~~~~3.2 Matter content and deeper unification?\\
~~~~~~~~~~3.3 Neutrino masses: evidence for a new very large mass scale?\\
~~~~~~~~~~3.4 Hierarchy problem in the Standard Model: hint for a new\\
~~~~~~~~~~~~~~~~~$\phantom{aaa}$ {\it low}  mass scale?\\
~~~~~~~~~~3.5 New low mass scale and precision electroweak data?
\item [4.]Supersymmetric extensions of the Standard Model\\
~~~~~~~~~~4.1 Precision electroweak data\\
~~~~~~~~~~4.2 The electroweak symmetry breaking\\
~~~~~~~~~~4.3 The mass of the lightest Higgs boson\\
~~~~~~~~~~4.4 Gauge coupling unification\\
~~~~~~~~~~4.5 Proton Decay
\item [5.]Summary
\end{enumerate}
\end{titlepage}

%\chapter I. {{\sc Review of the Standard Model (Electroweak Theory)}}
%\vspace{0.2cm}
\section{Introduction}

The Standard Model is a successful theory of interactions of quarks and 
leptons at energies up to about hundred GeV. Despite that success it is 
widely expected that there is physics beyond the Standard Model, with new 
characteristic mass scale(s), perhaps up to, ultimately, a string scale. 

The expectation is motivated by several fundamental questions that remain 
unanswered by the Standard Model. The most pressing one is better 
understanding of the mechanism of the electroweak symmetry breaking. The 
origin of flavour and of the pattern of  fermion masses and of CP violation 
also remain beyond its  scope. Moreover, we know  now  that the physics of 
the Standard Model cannot explain the baryon asymmetry in  the Universe. 
And on the top of all that come two recent strong experimental hints for 
physics beyond the Standard Model, that is very small neutrino masses and 
the presence of dark matter in the Universe. The list can be continued by 
including dark energy and inflation.

The Standard Model does not explain the scale of the electroweak symmetry 
breaking. It is a free parameter of the theory, taken from experiment.
Moreover, once we accept the point of view that the Standard Model is only an
effective ``low energy'' theory which is somehow cut-off at a mass scale $M$, 
and if $M\gg M_{W,Z}$, the electroweak symmetry breaking mechanism based on 
use of an  elementary Higgs field is unstable against quantum corrections 
(this is the so-called hierarchy problem).

Many different extensions of the Standard Model have been proposed to avoid  
the hierarchy problem and, more ambitiously, to calculate the scale of the  
electroweak breaking in terms of, hopefully, more fundamental parameters. 
Some extensions give the complete Standard Model, with one Higgs doublet, as 
their low energy approximation in the sense of the Appelquist - Carazzone 
decoupling and in some others the mechanism of the electroweak symmetry 
breaking cannot be decoupled from the bigger theory. The general idea is 
that the bigger theory has some characteristic mass scale $M$ only order of 
magnitude bigger than  $M_{W,Z}$, which  plays the role of a cut-off to the 
electroweak sector. All those extensions of the Standard Model 
have distinct experimental signatures. The experiments at the LHC will, 
hopefully, shed more light on the mechanism of the electroweak symmetry 
breaking and will support one of those (or still another one ?) directions.

One approach is based on low energy supersymmetry. The scale $M$ is 
identified with the mass scale of supersymmetric partners of the Standard 
Model particles. Supersymmetry is distinct in several very important points 
{}from all other proposed solutions to the hierarchy problem. First of all, 
it provides a general theoretical framework which allows to address many 
physical questions. Supersymmetric models, like  the Minimal Supersymmetric 
Standard Model or its simple extensions satisfy a very important criterion 
of ``perturbative calculability''. In particular, they are easily consistent 
with  the precision electroweak data. The Standard Model is their low energy 
approximation in the sense of the Appelquist - Carazzone decoupling, so most 
of the successful structure of the Standard Model is built into supersymmetric 
models. Unfortunately, there are also some troublesome exceptions: there are 
new potential sources of Flavour Changing Neutral Current (FCNC) transitions 
and of $CP$ violation, and baryon and lepton numbers are not automatically 
conserved by the renormalizable couplings. But even those problems can
at least be discussed in a concrete way. The quadratically 
divergent quantum corrections to the Higgs mass parameter (the origin of the 
hierarchy problem in the Standard Model) are absent in any order of 
perturbation theory. Therefore, the cut-off to a supersymmetric theory can 
be as high as the Planck scale and ``small'' scale of the electroweak  
breaking is still natural. But the hierarchy problem of the electroweak scale 
is solved at the price of a new hierarchy problem of the soft supersymmetry 
braking scale versus the Planck (string) scale. Spontaneous supersymmetry  
breaking and its transmission to the visible sector is a  difficult problem 
and a fully satisfactory mechanism has not yet been found. Again on the 
positive side, supersymmetry is not only consistent with Grand Unification 
of elementary forces but, in fact, makes it very 
successful. And, finally, supersymmetry is needed for string theory.

All other extensions of the electroweak theory proposed as solutions to the 
hierarchy problem rely on an onset of some kind of strong dynamics at energy 
scales not much higher than the electroweak scale. In some of them, like 
Higgless models with dynamical electroweak symmetry breaking or strong gravity
in large extra dimensions, the strong dynamics is simply a cut-off directly 
to the electroweak sector and appears already at ${\cal O}(1$~TeV). In  
models with the Higgs boson as a pseudo-Goldstone boson (e.g. Little Higgs 
models) and models with gauge fields present in extra dimensions the cut-off 
scale $M$ is identified with the characteristic scale of new perturbative 
physics, e.g. with the scale of breaking of some global symmetry or with the 
radii of extra dimensions. However, since those models are non-renormalizable 
and, moreover, in the bigger theory the quadratic divergences to the scalar 
mass parameter are absent typically only at one loop level, new physics itself 
has to be cut-off by some unknown strong dynamics at a scale one or two orders 
of magnitude higher than the $M$. Generally speaking, there is no 
Appelquist - Carazzone decoupling of new physics and the precision tests of 
such a version of the electroweak theory are not possible at the same level 
of accuracy as in the renormalizable Standard Model.

It is clear that models with early onset of strong dynamics cannot be easily, 
if at all, reconciled with Grand Unification. Also, they are very strongly 
constrained by precision electroweak data. There have been constructed
models that work but simple models are usually ruled out. Moreover, various 
aspects of flavour physics are often very obscure.

Spontaneous symmetry breaking in the condensed matter physics and in QCD is 
due to some collective effects. In supersymmetric models, such effects are 
presumably responsible for spontaneous breaking of supersymmetry and, in 
consequence, for the generation of soft mass terms. However, the electroweak 
symmetry breaking is driven by perturbative quantum corrections, generated 
by the large top quark Yukawa coupling, to the scalar potential. In the 
Little Higgs models, the Higgs boson is a Goldstone  boson of a bigger 
spontaneously broken global symmetry group. The Higgs potential 
needed for the electroweak symmetry breaking is also given by quantum 
corrections, with important contribution from the top quark Yukawa coupling.
Thus, one thing many models have in common  is that the electroweak symmetry   
is broken by perturbative quantum effects and linked to the large mass of 
the top quark.

At present, all extensions of the Standard Model remain speculative and none 
is fully satisfactory. Remembering the simplicity, economy and success of 
the Standard Model, one may wonder if in our search for its extensions 
shouldn't the Hipocrates principle {\it Primum non nocere} play more important 
role than it does. Indeed, various new ideas offer surprisingly low ratios 
of benefits to losses. It is, therefore, appropriate to begin by reviewing 
the basic structure of the Standard Model that underlies its success. It is 
likely that it  gives us important hints for the physics beyond.

\section {The Standard Model }

~~~~~{\bf 2.1 Basic structure}
\vskip0.2cm

The underlying principles  of the electroweak theory  are:
\begin{enumerate}
\item local $SU(2)_L \times U(1)_Y$ gauge symmetry and electroweak 
unification
\item spontaneous breaking of $SU(2)_L \times U(1)_Y$ gauge symmetry to 
 $U(1)_{EM}$, by the Higgs mechanism  with one Higgs doublet
\item matter content (chiral fermions) 
\item renormalizability
\end{enumerate}

Massless chiral fermions are the fundamental objects of matter: left-handed,
with helicity $\lambda=-1/2$, and right-handed, with helicity $\lambda=1/2$.
It is so because parity and charge conjugation are not the symmetries of our
world. The left-handed fermions carry different weak charges from the 
right-handed fermions. Chiral fermion fields are two-component (Weyl) spinors
(see e.g.~[1]):
\\
\begin{center}
\underline{$SU(2)_L$ doublets}
\end{center}
\begin{equation}
q_1\equiv
\left(
\begin{array}{c}
u\\
d
\end{array}
\right)~~~~~
q_2\equiv
\left(
\begin{array}{c}
c\\
s
\end{array}
\right)~~~~~
q_3\equiv
\left(
\begin{array}{c}
t\\
b
\end{array}
\right)
\label{1}
\end{equation}
\\
\begin{equation}
l_1\equiv
\left(
\begin{array}{c}
\nu_e\\
e
\end{array}
\right)~~~~~
l_2\equiv
\left(
\begin{array}{c}
\nu_{\mu}\\
\mu
\end{array}
\right)~~~~~
l_3\equiv
\left(
\begin{array}{c}
\nu_{\tau}\\
\tau
\end{array}
\right)
\label{2}
\end{equation}
with the electric charge and the hypercharge ($Y=Q-T_3$) assigned as below
\begin{center}
\begin{tabular}{c|cccc}
~&u &d & $\nu$ &e\\ \hline
Q &2/3 &-1/3 &0 &-1\\ \hline
Y &1/6 &1/6 &-1/2 &-1/2
\end{tabular}
\end{center}
These are left-handed chiral fields in the representation $(0,1/2)$ of the 
$SL(2,C)$, each  describing two massless degrees of 
freedom: a particle with the helicity $\lambda = -1/2$
and  its antiparticle with $\lambda = + 1/2$.
(The chiral fields can also be written  as four-component spinors  (see
e.g. \cite{POK}) but in the following we shall be using the Weyl notation). 
\\
\\
Right-handed fields  $[(1/2,0)$ of $SL(2,C)]$  in the same representations
of $SU(2)_L\times U(1)_Y$ as the left-handed fields (\ref{1}) and (\ref{2}) 
do not exist in Nature. Instead, we have

\begin{center}
\underline{$SU(2)_L$ singlets}
\end{center}

\[
\begin{array}{l}
u_R,~~~c_R,~~~t_R\\
d_R,~~~s_R,~~~b_R\\
e_R,~~~\mu_R,~~~\tau_R
\end{array}
\]
in $(1,+2/3),~~(1,-1/3)$ and $(1,-1)$ of $SU(2)_L\times U(1)_Y$, respectively.
These are right-handed chiral fields in the $(1/2,0)$ representation  of the
group $SL(2,C)$. For constructing a Lorentz invariant Lagrangian,
it is more convenient to take as fundamental fields only the left-handed
chiral fields. Thus, we introduce left-handed chiral fields, e.g
\begin{equation}
u^c,~~~c^c,~~~t^c
\label{3}
\end{equation}
in $(1,-2/3)$ of $SU(2)_L\times U(1)_Y$, such that
\begin{equation}
\bar{u^c}\equiv CPu^c(CP)^{-1}=u_R
\label{4}
\end{equation}
Indeed, $CP$ transformation results in the simultaneous change of chirality
and charges (representation $R\rightarrow R^*$ for internal symmetries).
Moreover, we see that the electric charge $Q=T_3+Y$ satisfies, e.g.
\begin{equation}
Q_{u^c}=-2/3=-Q_u
\label{5}
\end{equation}
and the two left-handed fields $u$ and $u^c$ become charge conjugate to each 
other when $U_{EM}(1)$ remains  the only  unbroken symmetry:
\begin{equation}
Cu^cC^{-1}=u
\label{6}
\end{equation}

We note that the matter chiral fields of the SM do not include a 
right-handed neutrino field $\nu_R$ in $(1,1)$ of $SU(2)_L\times U(1)_Y$ 
(such a charge assignment preserves
the relation $Q=T_3+Y$ ) or equivalently, a left-handed field
$\nu^c$ such that
\begin{equation}
\nu_R=CP\nu^c(CP)^{-1}
\label{7}
\end{equation}
but we can supplement the Standard Model with such a particle, if useful.

The breaking of the electroweak symmetry is generated by the potential of  
the Higgs doublet 
$ H= \left( 
\begin{array}{c}
H^+\\
H^0
\end{array}
\right)$ 
with the hypercharge $Y=+1/2$:
\begin{equation}
V=m^2H^\dagger H + {\frac {\lambda}{2}} (H^\dagger H)^2
\label{8}
\end{equation}
When $m^2<0$ is chosen, the Higgs doublet acquires the vacuum expectation 
value. Indeed, the minimum of the potential is for 
\begin{equation}
\langle H^\dagger H\rangle=-{m^2\over\lambda}\equiv{v^2\over2}
\label{9}
\end{equation}
By $SU_L(2)$ rotation we can always redefine the vacuum so that only the 
VEV  of the lower component of the Higgs doublet
is non-zero.  The $SU_L(2)\times U_Y(1)$ symmetry is then  broken down to 
$U^\prime (1)$ which is identified with
$U_{EM}(1)$ with  $Q=T^3 +Y$ because $(T^3 + Y)\left( 
\begin{array}{c}
0\\
v
\end{array}
\right)=0$.
The parameters $m$ and $\lambda$ are free parameters of the Standard Model. 
Equivalently, the scale of the electroweak symmetry breaking is not 
predicted by the theory  and must be taken from experiment.
\vskip0.3cm

{\bf 2.2 Fermion masses}
\vskip0.2cm

Higgs doublets (and only doublets) have $SU(2)_L\times U(1)_Y$ invariant  
renormalizable couplings to the chiral fermions of the Standard Model.
For the charged fermions, we can write down the following Yukawa couplings:
\begin{equation}
{\cal L}_{Yukawa}= -Y^{BA}_lH^*_il_{iA}e^c_B -Y_d^{BA}H^*_iq_{iA}d^c_B
-Y_u^{BA}\epsilon_{ij}H_iq_{iA}u^c_B+hc
\label{12}
\end{equation}
where $i$ is the $SU(2)_L$ index and $A,B$ are generation indices. We use the 
fact that the two-dim representation of $SU(2)$ is real and $i\tau_2H$
transforms as $H^\ast$, i.e. as $2^\ast(\equiv 2)$ of $SU(2)$. Therefore, 
$(i\tau_2Hq)=\epsilon_{ij}H_iq_j$ is also an invariant of $SU(2)$.
After spontaneous breaking of $SU(2)_L\times U(1)_Y$ to $U(1)_{EM}$ by the 
Higgs boson vacuum expectation value  $v$ we obtain the
Dirac masses
\begin{equation}
{\cal L}_{mass}=-v(Y_l^{BA}e_Ae^c_B+Y_d^{BA}d_Ad_B^c+Y_u^{BA}u_Au_B^c)+hc
\label{13}
\end{equation}
However, at the level of the full, $SU(2)_L\times U(1)_Y$ invariant theory, 
there is no renormalizable term that would give neutrino mass. It is so 
because $\nu^c$ is absent from the spectrum  of the SM. Thus, in the SM, 
neutrinos are massless.

The interactions (\ref{12},\ref{13}) are written in some "electroweak" 
basis defined 
by eigenvectors of the $SU(2)_L\times U(1)_Y$ symmetry group. In such a basis, 
both the fermion masses and the Yukawa couplings are in general non-diagonal 
in the flavour indices $(A,B)$. However, we can introduce another set of 
fields (say, primed fields) describing physical particles (mass eigenstates). 
The flavour of the primed fields is defined in the mass eigenstate basis. The 
two sets of fields are related to each other by unitary transformations:
\begin{equation}
\begin{array}{c}
\begin{array}{cc}
u= U_{L}u^{\prime }& d = D_{L}d^{\prime }\\
~\\
u^c = u^{\prime c} U_{R}^{\dagger}& d^c = d^{\prime C}D_{R}^{\dagger}\\
~\\
\end{array}\\
e = E_{L}e^{\prime }\\
~\\
e^c = e^{\prime c} E_{R}^{\dagger}
\end{array}
\label{14}
\end{equation}
which, of course, do not commute with the $SU(2)_L\times U(1)_Y$ gauge 
transformations and can be performed only after the 
spontaneous breakdown of the gauge symmetry. In eq.(\ref{14}), 
the fields $u,~d~, e$ denote three-dimensional vectors in the flavour space.

The transformations (\ref{14}) diagonalize the mass terms and the Yukawa 
couplings defined by (\ref{12}). After diagonalization
we can combine the chiral fields into Dirac fields. The weak currents
can be expressed in terms of the physical (mass eigenstates) fields:
\begin{eqnarray}
J^{-}_{\mu} &=&\sum_{A,B} \bar{u}^{\prime}_A 
\bar{\sigma}_\mu (V_{CKM})_{AB}d^{\prime}_B\nonumber \\
&+& \sum_A \bar{\nu}^{\prime}_A \bar{ \sigma}_\mu e^{\prime}_A
\label{15}
\end{eqnarray}
where the Cabibbo-Kobayashi-Maskawa matrix $V_{CKM} = U^{\dagger}_{L} D_L$. 
Note that the lepton current is diagonal in flavour (defined in the charged 
lepton mass eigenstate basis) because the \underline{massless} neutrino
field can be redefined by the transformation $\nu_A=E^{L}_{AB}\nu^\prime_B$ 
where $E^{L}_{AB}$ is the transformation diagonalizing the charged lepton 
mass matrix (see (\ref{14})). Thus, for the lepton current, 
$V_{CKM}=E^{L\dagger}E^L=1$.

It is important to remember that in the SM (with only one Higgs doublet) 
the Yukawa couplings to the physical Higgs boson (and, in fact, also the 
couplings to the $Z^0$) and the mass terms are diagonalized by the same 
unitary rotations. So they are flavour diagonal. The only source of flavour 
non-conservation resides in $V_{CKM}$. In particular, not only the global 
lepton number but also each flavour lepton number is separately conserved.
\vskip0.3cm

{\bf 2.3 Approximate custodial symmetry of the Standard Model and the 
precision electroweak data}\\
\vskip0.2cm

The Higgs sector of the SM is invariant under global $SO(4)$ symmetry acting
on four real components of the complex doublet. The group
$SO(4)\simeq SU(2)_L\times SU(2)_R$ and the Higgs doublet can be written as a 
$2\times2$ matrix $\Phi$
\begin{eqnarray}
\Phi=\left(\matrix{H^+&H^{0\ast}\cr H^0&-H^-}\right)~,
\end{eqnarray}
which transforms as $(\mathbf{2},\mathbf{2})$ of the latter 
group (whose first factor is just the gauged weak isospin group):
\begin{eqnarray}
\Phi\longrightarrow \Phi^\prime=U_L \Phi U_R~.
\end{eqnarray}
The vacuum expectation value of the Higgs field breaks $SU(2)_L\times SU(2)_R$
to its diagonal subgroup called ``custodial'' $SU(2)$ acting on the three
would-be Goldstone bosons $G^a$:
\begin{eqnarray}
H=\left(\matrix{H^+\cr H^0}\right)\longrightarrow 
{1\over\sqrt2}e^{iG^a\tau^a\over v}\left(\matrix{0\cr v+h^0}\right)
\end{eqnarray}
In the rest of the electroweak Lagrangian the $SU(2)_R$ subgroup and therefore
also the custodial $SU(2)$ symmetry is broken by the Yukawa interactions
and by the $U(1)_Y$ coupling. However, the custodial symmetry is still
seen at the tree level since by the Higgs mechanism it ensures that the 
gauge bosons $W^+$, $W^-$ and $W^0$ of  $SU(2)_L$ transform as a custodial
triplet. In consequence, the ratio of the strength of charged and neutral
current interaction at the tree level is equal to one
\begin{eqnarray}
\rho\equiv{M^2_W\over\cos^2\theta_WM^2_Z}=1~,
\end{eqnarray}
where $\cos^2\theta_W=g^2_2/(g^2_2+g^2_1)$.

This relation is consistent to a very good approximation with experimental
data but this is not the end of the success of the SM. Since the rest of the 
Lagrangian violates the custodial symmetry, there are quantum corrections
to the relation $\rho=1$. Since the global $SU(2)_L\times SU(2)_R$ symmetry
of the Higgs sector fixes the structure of counterterms in the scalar 
potential, quantum corrections to the relation $\rho=1$ must be finite!
In one-loop approximation one gets
\begin{eqnarray}
\Delta\rho\equiv\rho-1={3g^2_2\over64\pi^2}{m^2_t-m^2_b\over M^2_W}
-{g^2_1\over64\pi^2}{11\over3}\ln{M^2_h\over M^2_W}~.\label{eqn:drhoSM}
\end{eqnarray}
The first term is of the order of 1\% and, is in perfect agrrement with 
the precision electroweak data. Thus, fits to the data give 
$M_h\sim{\cal O}(M_W)$ although logarithmic dependence of $\Delta\rho$ on
$M_h$ does not allow for precise determination of this mass. We shall discuss 
later on the importance of the Higgs particle mass for various extensions of 
the SM.

We conclude that the approximate custodial symmetry of the SM is in fantastic
agreement with experimental data. Any alternative mechanism of the electroweak 
symmetry breaking or any extension of the SM must not violate the custodial 
symmetry of the electroweak interactions. Furthermore, we see that the 
renormalizable SM with one Higgs doublet, has very strong predictive power 
which allows for its precision tests at the level of one per mille.
 
\vskip0.3cm

{\bf 2.4 GIM mechanism  and the suppression of FCNC and CP violating 
transitions}
\vskip0.2cm

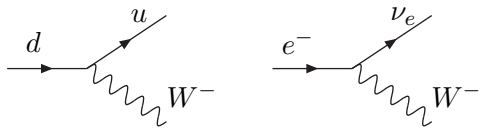
\begin{figure}[htbp]
\begin{center}
%\begin{tabular}{lp{280\unitlength}}
\begin{picture}(210,60)(0,0)
\ArrowLine(20,30)(50,30)
\ArrowLine(50,30)(80,50)
\Photon(50,30)(80,10){3}{5}
\Text(30,40)[]{$d$}
\Text(70,50)[]{$u$}
\Text(90,20)[]{$W^-$}
\ArrowLine(120,30)(150,30)
\ArrowLine(150,30)(180,50)
\Photon(150,30)(180,10){3}{5}
\Text(130,40)[]{$e^-$}
\Text(170,50)[]{$\nu_e$}
\Text(190,20)[]{$W^-$}
\end{picture}
\end{center}
\caption{Charged current transitions in the Standard Model.}
\label{fig:CCelementary}
%\end{tabular}
\end{figure}

\begin{figure}[htbp]
\begin{center}
%\begin{tabular}{lp{280\unitlength}}
\begin{picture}(310,60)(0,0)
\ArrowLine(20,30)(50,30)
\ArrowLine(50,30)(80,50)
\Photon(50,30)(80,10){3}{5}
\Text(30,40)[]{$d$}
\Text(70,50)[]{$s$}
\Text(90,20)[]{$Z^0$}
\ArrowLine(120,30)(150,30)
\ArrowLine(150,30)(180,50)
\Photon(150,30)(180,10){3}{5}
\Text(130,40)[]{$\mu^-$}
\Text(170,50)[]{$e^-$}
\Text(190,20)[]{$Z^0$}
\ArrowLine(220,30)(250,30)
\ArrowLine(250,30)(280,50)
\DashLine(250,30)(280,10){3}
\Text(230,40)[]{$d$}
\Text(270,50)[]{$s$}
\Text(290,20)[]{$h^0$}
\end{picture}
\end{center}
\caption{Neutral current transitions absent at the tree level in the SM.}
\label{fig:NCelementary}
%\end{tabular}
\end{figure}
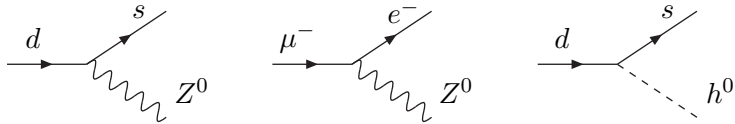

It is well established experimentally that the amplitudes for processes such 
as e.g. $K^0(d\bar s)$-$\bar K^0(\bar ds)$ mixing caused by electroweak
interactions, $\langle K^0|{\cal H}_{\rm weak}|\bar K^0\rangle$, are strongly 
suppressed in comparison with the amplitudes for the charged current 
transitions like $n\rightarrow pe^-\bar\nu_e$ or
$\mu^-\rightarrow e^-\bar\nu_e\nu_\mu$. A good measure of the mixing is
the mass difference between neutral kaon mass eigenstates 
$|K^0_L\rangle={1\over\sqrt2}(|K^0\rangle-|\bar K^0\rangle)$ and
$|K^0_S\rangle={1\over\sqrt2}(|K^0\rangle+|\bar K^0\rangle)$ (we neglect
here even smaller CP violation): $\Delta M_K=3\times10^{-12}$~MeV and is
suppressed by factor $10^6$ compared to what one could expect for a generic
electroweak transition. This fact finds a very elegant explanation in the 
Standard Model. The charged current transitions shown in 
Fig.~\ref{fig:CCelementary} are present at the tree level
whereas it follows from the structure of the theory that the diagrams
shown in Fig.~\ref{fig:NCelementary}
are absent. In the diagrams the quark fields are of course mass
eigenstate fields and the couplings are obtained by rotating from an
electroweak basis, in which the theory is formulated, to the mass eigenstate
basis, in which the quark flavour is defined.

It is obvious that the $Z^0$ couplings are flavour diagonal. More 
interesting is the absence in the SM of the scalar flavour changing neutral
currents. This result follows from the fact that in the SM there is only one
Higgs doublet. Because of that, diagonalizing the fermion mass matrices one 
simultaneously obtains also diagonal Yukawa couplings to the physical Higgs
boson. In models with more Higgs doublets, additional discrete symmetry has
to be imposed to ensure that only one Higgs doublet couples to the quarks
of the same charge. In the minimal supersymmetric model
two doublets are needed for supersymmetric theory but (by the holomorphicity
of the superpotential) only one Higgs doublet can couple to
the same charge quarks.

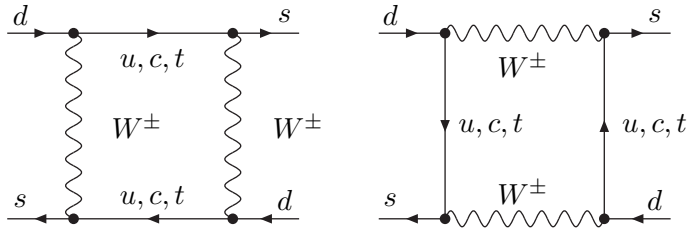
\begin{figure}[htbp]
\begin{center}
%\begin{tabular}{lp{280\unitlength}}
\begin{picture}(250,100)(0,0)
\ArrowLine(0,80)(25,80)
\ArrowLine(25,80)(85,80)
\ArrowLine(85,80)(110,80)
\ArrowLine(110,10)(85,10)
\ArrowLine(85,10)(25,10)
\ArrowLine(25,10)(0,10)
\Vertex(25,80){2}
\Vertex(85,80){2}
\Vertex(25,10){2}
\Vertex(85,10){2}
\Photon(25,80)(25,10){3}{7}
\Photon(85,80)(85,10){3}{7}
\Text(49,45)[]{$W^\pm$}
\Text(110,45)[]{$W^\pm$}
\Text(5,87)[]{$d$}
\Text(105,87)[]{$s$}
\Text(05,17)[]{$s$}
\Text(105,17)[]{$d$}
\Text(55,70)[]{$u,c,t$}
\Text(55,18)[]{$u,c,t$}
\ArrowLine(140,80)(165,80)
\ArrowLine(225,80)(250,80)
\ArrowLine(250,10)(225,10)
\ArrowLine(165,10)(140,10)
\Vertex(165,80){2}
\Vertex(225,80){2}
\Vertex(225,10){2}
\Vertex(165,10){2}
\ArrowLine(165,80)(165,10)
\ArrowLine(225,10)(225,80)
\Photon(165,80)(225,80){3}{7}
\Photon(225,10)(165,10){3}{7}
\Text(183,45)[]{$u,c,t$}
\Text(245,45)[]{$u,c,t$}
\Text(145,87)[]{$d$}
\Text(245,87)[]{$s$}
\Text(145,17)[]{$s$}
\Text(245,17)[]{$d$}
\Text(195,68)[]{$W^\pm$}
\Text(195,20)[]{$W^\pm$}
\end{picture}
\end{center}
\caption{Leading SM contribution to $K^0$-$\bar K^0$ mixing.}
\label{fig:box}
%\end{tabular}
\end{figure}

The absence of flavour non-diagonal neutral currents at the tree level is not
sufficient to account for  the observed suppression of processes like
kaon mixing or $b\rightarrow s\gamma$. For example the 1-loop diagrams shown 
in Fig.~\ref{fig:box} generate the $K^0(d\bar s)$-$\bar K^0(\bar ds)$
transitions and the coefficient $C$ in the effective Lagrangian 
\begin{eqnarray}
{\cal L}_{\rm eff}=C~(\bar s_L\gamma_\mu d_L)(\bar s_L\gamma_\mu d_L)
\label{eqn:effLKK}
\end{eqnarray}
describing their contribution (in the limit of external quark momenta
small compared to $M_W$) is generically of order 
$C\sim{\alpha^2\over M_W^2}\sim\alpha ~G_F$. However, in the SM the sum 
of all such contributions is suppressed by a factor $\sim10^{-4}$ due 
to the so-called (generalized to 3 generations of quarks) 
Glashow-Illiopoulos-Maiani mechanism.
The coefficient $C$ generated by diagrams of Fig.~\ref{fig:box}
(and the diagrams in which one or both $W^\pm$ are replaced  by the
unphysical would-be Goldstone bosons $G^\pm$) is finite and has dimension
mass$^{-2}$. The whole effective Lagrangian can be written as
\begin{eqnarray}
{\cal L}_{\rm eff}&=&
{1\over 2}\left({g_2\over\sqrt2}\right)^4\sum_{i,j=u,c,t}V^{\star}_{is} 
V_{id} V^{\star}_{js} V_{jd}\label{box1}\\
&\times&\int {d^4 q\over(2\pi)^4} 
{[\overline\Psi_s\gamma_\mu P_L(\not{q}+m_{q_i})\gamma_\nu P_L \Psi_d]
 [\overline\Psi_s\gamma^\nu P_L(\not{q}+m_{q_j})\gamma^\mu P_L \Psi_d]\over
(q^2 - m_{q_i}^2) (q^2-m_{q_j}^2) (q^2 - M_W^2)^2}\nonumber
\end{eqnarray}
The top quark contribution is suppressed by the smallness of the product
$V_{ts}^\ast V_{td}$. The rest contributes to the coefficient $C$ in 
(\ref{eqn:effLKK}) 
\begin{eqnarray}
A_{uc} &\sim& \left({g_2\over\sqrt2}\right)^4{1\over M^2_W} 
\sum_{i,j=u,c}V^{\star}_{is} V_{id}V^\star_{js} V_{jd}
\left[1 + {\cal O}\left( 
{m^2_{q_i}\over M^2_W}, {m^2_{q_j}\over M^2_W}
\right)\right]
\phantom{a}\nonumber\\
&\sim&\alpha ~G_F\left\{\left( V^\star_{ts} V_{td} \right)^2 +  
{\cal O} \left( \sum_{i,j=u,c} V^{\star}_{is} V_{id} V^{\star}_{js} V_{jd} 
{m^2_{q}\over M^2_W} \right) \right\}
\end{eqnarray}
where in the last line, unitarity of the CKM matrix has been used: 
$V^{\star}_{us}V_{ud}+V^{\star}_{cs}V_{cd}=-V^{\star}_{ts} V_{td}$.

 From this example it is clear that for the empirical pattern of quark masses 
and mixing angles there is strong suppression of FCNC in the SM. It is much 
stronger than ``naturally'' expected. 

The predictions of the SM for the FCNC transitions are in very good agreement
with experimental data. This is also true for CP violation. The only source
of CP violation in the SM is the phase of the CKM matrix. As a result, the 
effects of CP violation in the kaon system, in which they were first observed, 
are proportional to the masses of the light quarks and small CKM mixing angles 
and hence very small (this is not so for the $B$-meson systems in which CP 
violation is probed by present experiments). 

The strong suppression of the FCNC and CP violating transitions, so nicely
consistent with the SM is a big challenge for any of its extension. This is
easy to understand on a qualitative basis. Let us suppose that new physics
contributes to such transitions at 1-loop level (any contribution at the
tree level would be a total disaster!) with the couplings of order of the
strong coupling constant $\alpha_s\approx0.12$ and with the scale $M$
of the particle masses in the loop. Then
\begin{eqnarray}
\Delta C \sim{\alpha_s^2\over M^2}={\alpha^2\over M_W^2}
\left({\alpha_s\over\alpha}\right)^2
\left({M_W\over M}\right)^2
\end{eqnarray}
Thus, for such contributions to be comparable or smaller than the SM one, 
the new physics scale $M$ has to be higher than $10^3M_W\sim100$~TeV. If 
the new couplings are of the order of $\alpha$ then we get $M\simgt10$ TeV. 
If the scale of new physics extension of the SM is below these limits, the 
new physics must somehow control the flavour effects!
\vskip0.3cm

{\bf 2.5 Baryon and lepton number conservation}
\vskip0.2cm

The principles (1) - (4) \underline{imply} global $U(1)$ symmetries of 
the theory: baryon and lepton number conservation $\Delta B=\Delta L=0$. 
In fact, for leptons the implication is even stronger, namely 
$U_e(1)\times U_\mu(1)\times U_\tau(1)$ is a global symmetry of the 
electroweak Lagrangian and the lepton flavour numbers are separately 
conserved: $\Delta L_e=\Delta L_\mu=\Delta L_\tau$=0. For quarks, quark mixing 
explicitly breaks quark flavour $U(1)$'s and only the total baryon number is 
conserved. 

\begin{figure}[htbp]
\begin{center}
%\begin{tabular}{lp{280\unitlength}}
\begin{picture}(230,80)(0,0)
\ArrowLine(20,40)(70,70)
\ArrowLine(70,70)(70,10)
\ArrowLine(70,10)(20,40)
\Line(10,41)(20,41)
\Line(10,39)(20,39)
\Vertex(20,40){2}
\Text(-5,40)[]{$j^{L,B}_\mu$}
\Line(70,71)(80,71)
\Line(70,69)(80,69)
\Vertex(70,70){2}
\Text(95,70)[]{$j^a_{\nu L}$}
\Line(70,11)(80,11)
\Line(70,9)(80,9)
\Vertex(70,10){2}
\Text(95,10)[]{$j^b_{\kappa L}$}
\Text(80,40)[]{$f_L$}
\ArrowLine(160,40)(210,10)
\ArrowLine(210,10)(210,70)
\ArrowLine(210,70)(160,40)
\Line(150,41)(160,41)
\Line(150,39)(160,39)
\Vertex(160,40){2}
\Text(135,40)[]{$j^{L,B}_\mu$}
\Line(210,71)(220,71)
\Line(210,69)(220,69)
\Vertex(210,70){2}
\Text(235,70)[]{$j^a_{\nu L}$}
\Line(210,11)(220,11)
\Line(210,9)(220,9)
\Vertex(210,10){2}
\Text(235,10)[]{$j^b_{\kappa L}$}
\Text(220,40)[]{$f_L$}
\end{picture}
\end{center}
\caption{Anomalies in the $B$ and $L$ currents in the SM.}
\label{fig:anomaly}
%\end{tabular}
\end{figure}
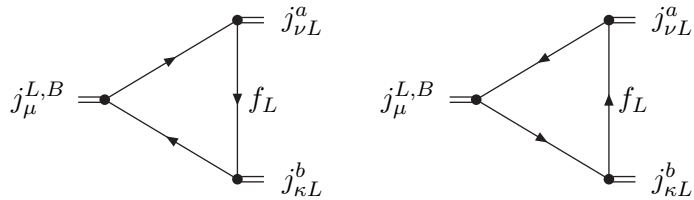

The conservation of the baryon and lepton numbers by the renormalizable 
couplings of the Standard Model is beautifully consistent with experimental 
limits on the life time of the proton , $\tau_p\simgt10^{33}$ years, and on 
the branching ratios for the lepton flavour violating decays, e.g. 
$BR(\mu\rightarrow e\gamma)\simlt10^{-11}$. Proton decay and lepton flavour 
violating decays occur, if at all, many orders of magnitude less frequently 
than generic electroweak processes. Actually, in the Standard Model those 
conservation laws are violated by chiral anomaly. The diagrams shown in 
Fig.~\ref{fig:anomaly} where $j^a_{\mu L}$'s are $SU(2)_L$ gauge currents 
and $j_\mu^{L,B}$ is the baryon or lepton current of the
$U(1)$ global symmetries gives (insisting on the conservation of the gauge 
currents  [1])
\begin{eqnarray}
&&\partial^\mu j^B_\mu\propto({\rm Tr}B)
\sum_{a=1}^3W^a_{\mu\nu}{\tilde W^{a\mu\nu}}
\nonumber\\
&&\partial^\mu j_\mu^L\propto({\rm Tr}L)~
\sum_{a=1}^3W_{\mu\nu}^a{\tilde W^{a\mu\nu}}
\end{eqnarray}

Only $SU_L(2)$ doublets contribute to the traces, so they do not vanish, 
and $W^a_{\mu\nu}$ is the $SU(2)_L$ field strength. Non-perturbative effects 
give, in general a non-zero condensate $W^a_{\mu\nu}{\tilde W^{\mu\nu}}$ 
(topological baryon and lepton number non-conservation) but the effect is 
totally negligible at zero temperature. At non-zero temperature, the 
topological baryon and lepton number non-conservation is enhanced and can 
play important physical role because the $(B-L)$ current is anomaly free:
$\partial^\mu j^{B-L}_\mu\propto{\rm Tr}(B-L)=0$ and the quantum number
$B-L$ is conserved. Thus in the presence of 
some hypothetical perturbative lepton number and $CP$ violation, topological 
effects may convert leptogenesis into baryogenesis.

\begin{figure}[htbp]
\begin{center}
%\begin{tabular}{lp{280\unitlength}}
\begin{picture}(230,80)(0,0)
\ArrowLine(20,40)(70,70)
\ArrowLine(70,70)(70,10)
\ArrowLine(70,10)(20,40)
\Photon(10,40)(20,40){2}{3}
\Vertex(20,40){2}
\Text(-5,40)[]{$j^{B-L}_\mu$}
\Photon(70,70)(80,70){2}{3}
\Vertex(70,70){2}
\Text(95,70)[]{$j^{B-L}_{\nu L}$}
\Photon(70,10)(80,10){2}{3}
\Vertex(70,10){2}
\Text(95,10)[]{$j^{B-L}_{\kappa L}$}
\Text(80,40)[]{$f$}
\ArrowLine(160,40)(210,10)
\ArrowLine(210,10)(210,70)
\ArrowLine(210,70)(160,40)
\Photon(150,40)(160,40){2}{3}
\Vertex(160,40){2}
\Text(135,40)[]{$j^{B-L}_\mu$}
\Photon(210,70)(220,70){2}{3}
\Vertex(210,70){2}
\Text(235,70)[]{$j^{B-L}_{\nu L}$}
\Photon(210,10)(220,10){2}{3}
\Vertex(210,10){2}
\Text(235,10)[]{$j^{B-L}_{\kappa L}$}
\Text(220,40)[]{$f$}
\end{picture}
\end{center}
\caption{$[U(1)_{B-L}]^3$ anomaly.}
\label{fig:BLtriangle}
%\end{tabular}
\end{figure}
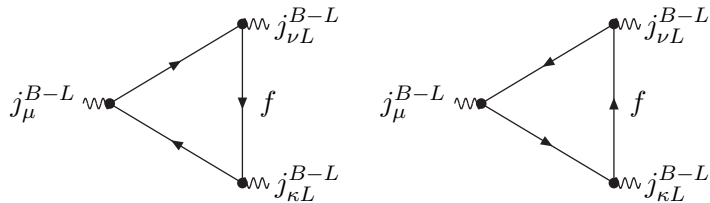

Incidentally, with the right-handed neutrino included in the spectrum, the 
diagrams in Fig.~\ref{fig:BLtriangle} do not give any anomaly, neither, and 
the $(B-L)$ symmetry can, therefore, be gauged.

\section {Hints from the Standard Model for its extensions}

~~~~~{\bf 3.1 Is the effective low energy electroweak theory indeed the 
      ~~~~~renormalizable Standard Model?}\\

In the construction of the SM Lagrangian we have been, so far, guided by
its renormalizability. Accepting the fact that the SM is only an effective
theory one may wonder, however, how important is its renormalizability.
Unitarity and symmetries are  certainly more fundamental requirements and
indeed e.g. the physics of pions is described by a non-renormalizable effective
low energy theory (non-linear $\sigma$-model). It is, therefore, useful to 
recall the main differences between the two classes of quantum field theories.

In a renormalizable theory its cut-off can be taken to infinity and the whole
UV sensitivity is hidden in a finite number of free parameters, the same
at any order of perturbation expansion. Calculations with arbitrary precision
are, therefore, possible with a fixed number of parameters whose values
can be determined from the experimental data. If some theory gives as its
low energy approximation a renormalizable theory, then according to the
Appelquist-Carazzone decoupling theorem, the effects of heavy degrees of
freedom characterized by a mass scale $M$ show up only as corrections
in the form of higher dimensional operators allowed by the symmetries 
of the renormalizable theory:
\begin{eqnarray}
{\cal L}_{\rm eff}={\cal L}_{\rm renormalizable}
+{\cal O}\left({1\over M^n}\right)~O_{4+n}
\end{eqnarray}
This is a window to new physics (if such corrections are needed by experiment)
even if we do not know the theory at the scale $M$.

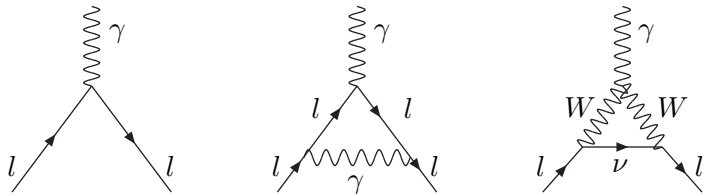
\begin{figure}[htbp]
\begin{center}
%\begin{tabular}{lp{280\unitlength}}
\begin{picture}(300,80)(0,0)
\Photon(50,70)(50,40){3}{6}
\ArrowLine(20,0)(50,40)
\ArrowLine(50,40)(80,0)
\Text(80,10)[]{$l$}
\Text(20,10)[]{$l$}
\Text(60,60)[]{$\gamma$}
\ArrowLine(120,0)(135,20)
\ArrowLine(135,20)(150,40)
\ArrowLine(150,40)(165,20)
\ArrowLine(165,20)(180,0)
\Photon(150,70)(150,40){3}{6}
\Photon(130,13)(170,13){3}{6}
\Text(180,10)[]{$l$}
\Text(120,10)[]{$l$}
\Text(150,3)[]{$\gamma$}
\Text(135,32)[]{$l$}
\Text(170,32)[]{$l$}
\Text(160,60)[]{$\gamma$}
\ArrowLine(220,0)(235,17)
\ArrowLine(235,17)(265,17)
\ArrowLine(265,17)(280,0)
\Photon(235,17)(250,40){3}{6}
\Photon(250,40)(265,17){3}{6}
\Photon(250,70)(250,40){3}{6}
\Text(280,10)[]{$l$}
\Text(220,10)[]{$l$}
\Text(250,10)[]{$\nu$}
\Text(235,32)[]{$W$}
\Text(270,32)[]{$W$}
\Text(260,60)[]{$\gamma$}
\end{picture}
\end{center}
\caption{One loop contributions to the anomalous lepton magnetic moment 
in the SM.}
\label{fig:gminusdwa}
%\end{tabular}
\end{figure}

It is worth putting the view at the Standard Model as an effective 
low energy theory into the better known perspective. We know now that Quantum 
Electrodynamics (QED) is a renormalizable theory and at the same time it is 
the low energy approximation to the electroweak theory. Its renormalizability 
means calculability with arbitrary precision. But it is only an effective 
theory so we know that its predictions disagree with experiment at the level 
$\sim {\cal{O}} (E/M_{\mbox{w}})$, where the energy $E$ is the characteristic 
energy for a given process. For example, let us have a look at the lepton 
magnetic moment. It gets contributions from the  diagrams depicted in 
Fig.~\ref{fig:gminusdwa}.
Thus, for the nonrelativistic effective interaction with the magnetic 
field we get
\begin{equation}
{\cal H}_{\rm eff}={e\over2m_l} 
\mbox{\boldmath$\sigma$}\cdot\mathbf{B}
(1+ {\alpha\over2\pi} 
+ {\cal O}\left (\alpha{m^2_l\over M^2_W}\right) + \ldots)
\label{27}
\end{equation}
where the role of the energy scale is played by the lepton mass $m_l$. The 
"weak" correction is calculable in the full electroweak theory, but at the 
level of QED as an effective theory it has to be added as a new 
non-renormalizable (but $U(1)_{EM}$ invariant) interaction
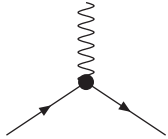
\begin{figure}[htbp]
\begin{center}
%\begin{tabular}{lp{280\unitlength}}
\begin{picture}(300,70)(0,0)
\ArrowLine(120,0)(150,20)
\ArrowLine(150,20)(180,0)
\Photon(150,50)(150,20){3}{6}
\Vertex(150,20){3}
\end{picture}
\end{center}
\caption{Effective photon-lepton vertex.}
\label{fig:effint}
%\end{tabular}
\end{figure}
\begin{equation}
{\cal L}_{\rm eff}={m_l\over M^2}\bar{\psi}\sigma_{\mu\nu}\psi 
F^{\mu\nu}\phantom{aaaaaaa} \mbox{(dim 5)}
\label{28}
\end{equation}
This would have been a way to discover weak interactions (and to measure the 
weak scale) in purely electromagnetic processes: we extend QED to a 
non-renormalizable theory by adding higher dimension operators and look for 
their experimental manifestation in purely electromagnetic processes once the 
experimental precision is high enough. Luckily enough for us, effective QED 
may also contain other than (\ref{28}) non-renormalizable corrections, 
$U(1)_{EM}$ invariant but violating the conservation of quantum numbers
that are accidentally conserved in QED, for instance flavour. Such corrections 
manifest themselves as different type of interactions - weak interactions - 
and were easy to discover experimentally. Similarly, among many possible
non-renormalizable corrections to the SM which respect the 
$SU(2)_L\times U(1)_Y$ gauge symmetry there are such that violate e.g. the
lepton and/or baryon number conservation or give Majorana masses to neutrinos.
We shall discuss them in the following.

In a non-renormalizable theory one either has to keep explicit logarithmic 
cut-off dependence (${1\over16\pi^2}\ln{\Lambda\over\mu}$ where $\mu$ is 
some low energy scale) or the number of counterterms (i.e. the number of 
free parameters of the theory) must increase at each order of perturbation 
expansion. The value of the cut-off $\Lambda$ is dictated by the consistency 
between one-loop calculations and the contribution of the higher dimensional 
operators. Typically, the theory becomes strongly interacting
above the cut-off scale.

A physically important non-renormalizable effective theory is the theory of
pions \cite{WE,POK}. The pions are pseudo-Goldstone bosons of the 
(approximate) global 
chiral symmetry $SU(2)_L\times SU(2)_R$ of strong interactions which is 
spontaneously broken down to $SU(2)_V$ of isospin. The physics of the light 
degrees of freedom (pions) is described by a non-linear $\sigma$-model. The 
chiral symmetry $SU(2)_L\times SU(2)_R$, non-linearly realized on the pion 
fields, requires non-renormalizable interaction. The lowest dimension one is
\begin{eqnarray}
{\cal L}_{\rm pions}=f^2_\pi{\rm Tr}(\partial_\mu U^\dagger\partial^\mu U)
\end{eqnarray}
where the fields
\begin{eqnarray}
U=e^{i\pi^a\tau^a/f_\pi}
\end{eqnarray}
transform under  $SU(2)_L\times SU(2)_R$ linearly:
$U\rightarrow V_L U V_R^\dagger$. The constanf $f_\pi$ is the pion decay
constant.
The chiral symmetry cannot be reconciled with an effective renormalizable
theory of pions and the Appelquist-Carazzone decoupling does not work.

The important question is: is the true low energy approximation to the more 
fundamental theory which explains the mechanism of electroweak symmetry 
breaking the renormalizable SM (like in supersymmetric extensions of the 
SM) or non-renormalizable electroweak theory (like in higgsless models 
with dynamical electroweak symmetry breaking and in models with the Higgs 
boson as a pseudo-Goldstone boson of some spontaneously broken bigger
global symmetry)?

The predictive power and the phenomenological success of the SM suggests the 
first case. On the other hand, one may argue that the second option would 
more resemble spontaneous symmetry breaking in the condensed matter physics 
and in strong interactions.
\vskip0.3cm

~~~~~{\bf 3.2 Matter content and deeper unification?}
\vskip0.2cm

There are two striking aspects of the matter spectrum in the Standard Model. 
One is the chiral anomaly cancellation \cite{WE,POK}, which is necessary for 
a unitary (and 
renormalizable) theory, and occurs thanks to certain  conspiracy between
quarks and leptons suggesting a deeper link between them. The potential source 
of chiral anomalies in the Standard Model are the triangle diagrams
like the ones shown in Figs.~\ref{fig:anomaly} in which now the external
lines correspond to all possible triplets of currents coupled to the three
types of gauge fields in the electroweak theory: $U(1)_Y$ gauge field
$B_\mu$, $SU(2)_L$ gauge fields $W_\mu^a$ ($a=1,2,3$) and/or $SU(3)_C$ gauge 
fields $A_\mu^a$ ($a=1,\dots,8$) and internal fermion lines correspond to all 
chiral fermions in the theory. Most of the anomaly coefficients
vanish due to the group structure. The most interesting ones are the anomalies
with one $U(1)_Y$ current and two $SU(2)_L$ currents and the one with three
$U(1)_Y$ currents. They vanish
(a necessary condition for consistency of the electroweak theory) provided
\begin{equation}
\sum Q_i = 0~,\label{eqn:anomalycond}
\end{equation}
(where $Q_i$'s are the electric charges of the fields) separately for doublets 
and singlets of $SU(2)_L$. 
Incidently, the same condition is sufficient for vanishing of the mixed 
$U(1)_Y$-gravitational anomaly given by the diagrams like those shown in 
Figs.~\ref{fig:anomaly} but now with two currents corresponding to the
energy-momentum tensors and the third one to the $U(1)_Y$ current. The 
condition (\ref{eqn:anomalycond}) is satisfied in the SM because quark 
and lepton contributions cancel each other.

The second striking feature of the matter spectrum in the Standard Model is 
that it fits into simple representations of the $SU(5)$ and $SO(10)$ groups
\cite{LAN}. Indeed, we have, for $SU(5)$
\begin{eqnarray}
\mathbf{5}^\ast&=&\left(\matrix{\nu_e\cr e^-}\right)_L , ~d_L^c\nonumber\\
\mathbf{10}&=&\left(\matrix{u\cr d_L}\right), ~u_L^c, ~e_L^c\nonumber\\
\mathbf{1}&=&\nu_L^c
%\nonumber\\
%&\phantom{a}&\phantom{aaaaaa}
\phantom{aaa}
{\rm (if ~the ~right-handed ~neutrino ~is ~added ~to ~the ~spectrum)}\nonumber
\end{eqnarray}
and, for $SO(10)$,
\begin{eqnarray}
\mathbf{16} =\mathbf{5}^\ast +\mathbf{10} +\mathbf{1}\nonumber
\end{eqnarray}
%~~~~~~~~~~~~~~~~~~~~$ 16 =5* +10 +1 $
The assignment of fermions to the $SU(5)$ representations fixes the 
normalization of the $U(1)_Y$ generator:
\begin{equation}
Q=T_3+Y=L^{11}+\sqrt{5\over3}L^{12}~,\label{eqn:so10gen}
\end{equation}
where $L^{ij}$ are the $SU(5)$ generators satisfying the normalization 
condition $[L^{ij},~L^{kl}]={1\over2}\delta^{ik}\delta^{jl}$.

Both facts, the anomaly cancellation  and the pattern of fermion spectrum, 
strongly suggest some kind of quark and lepton  unification, at least at 
some very deep level, with some big group and some mechanism of its breaking. 
In addition, in line with the above conclusion is a  well known fact that, 
with normalization given by eq~(\ref{eqn:so10gen}), the running gauge 
couplings of the Standard Model approach each other at high scale of order 
$10^{13}$~GeV. Although unification  of the gauge couplings in the Standard 
Model is only very approximate, it is nevertheless a remarkable fact that 
the strength of strong and electroweak interactions become comparable at 
certain energy scale.
\vskip0.3cm

{\bf 3.3 Neutrino masses: evidence for new very high mass scale?}
\vskip0.2cm

There is at present strong experimental evidence for neutrino oscillations 
whose most obvious and most natural explanation is that neutrinos are massive 
and their mass eigenstates are different from the weak interaction eigenstates.
This is the first experimental evidence for physics beyond the Standard Model.

The smallness of the neutrino masses can be easily understood as due to the 
presence of a new very high mass scale $M$. If it makes sense to rely on the 
Appelquist-Carazzone  decoupling theorem then the mass scale 
$M$ would manifest itself via higher dimension operators. 

Neutrino mass terms may appear  as dimension
five  operator
\begin{equation}
\frac{1}{M}(Hl_A)\lambda_{AB}(Hl_B)
\label{16}
\end{equation}
where we use the following  notation: $(Hl)\equiv \epsilon_{ij}
H^il^j$ denotes $SU(2)_L$ contraction and $ll\equiv \epsilon^{\alpha\beta}
l_{\alpha} l_{\beta}$ denote Lorentz contraction. After spontaneous SM 
gauge symmetry breaking by the Higgs boson VEV the operator (\ref{16}) gives 
indeed a Majorana mass matrix for  neutrinos:
\begin{equation}
{\cal L}_{\nu~\rm mass}=-m_{AB}~\nu_A\nu_B+{\rm H.c.}~,\phantom{aaaaaaa}
m_{AB}={v^2\over M}\lambda_{AB}
\label{17}
\end{equation}
Small neutrino masses are obtained for big value of $M$, with the 
constants $\lambda_{AB}\sim {\cal O}(1)$. This is called a see-saw mechanism.

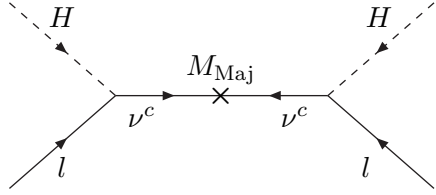
\begin{figure}[htbp]
\begin{center}
\begin{picture}(180,90)(0,0)
\ArrowLine(10,10)(50,45)
\ArrowLine(170,10)(130,45)
\ArrowLine(50,45)(90,45)
\ArrowLine(130,45)(90,45)
\DashArrowLine(10,80)(50,45){3}
\DashArrowLine(170,80)(130,45){3}
\Text(90,45)[]{$\mbox{\boldmath$\times$}$}
\Text(30,18)[]{$l$}
\Text(145,18)[]{$l$}
\Text(60,37)[]{$\nu^c$}
\Text(118,37)[]{$\nu^c$}
\Text(30,75)[]{$H$}
\Text(150,75)[]{$H$}
\Text(90,55)[]{$M_{\rm Maj}$}
\end{picture}
\end{center}
\caption{Diagram generating the dimension 5 operator.}
\label{fig:dim5op_gen}
\end{figure}

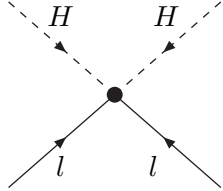
\begin{figure}[htbp]
\begin{center}
\begin{picture}(180,90)(0,0)
\ArrowLine(50,10)(90,45)
\ArrowLine(130,10)(90,45)
\DashArrowLine(50,80)(90,45){3}
\DashArrowLine(130,80)(90,45){3}
\Vertex(90,45){3}
\Text(70,18)[]{$l$}
\Text(105,18)[]{$l$}
\Text(70,75)[]{$H$}
\Text(110,75)[]{$H$}
\end{picture}
\end{center}
\caption{Effective dimension 5 operator.}
\label{fig:dim5op}
\end{figure}

A possible and, in fact quite elegant, origin of the mass scale $M$ would 
be the existence of another left-handed particle
$\nu^c$, a singlet of $SU(2)_L\times U(1)_Y$, i.e. a field such that
\begin{equation}
CP\nu^c(CP)^{-1}\equiv \nu_R
\label{29}
\end{equation}
with a Majorana mass term 
\begin{equation}
{\cal L}_{\rm Majorana} = M^{AB}_{\rm Maj}~\nu^c_A \nu^c_BB + {\rm H.c.}
\label{30}
\end{equation}
It can be interpreted as a right-handed neutrino field.
Moreover, we can construct Yukawa interactions
\begin{equation}
\epsilon_{ij}H_i\nu^c_BY^{BA}_{\nu}l^A_j+{\rm H.c.}
\label{31}
\end{equation}
with a new set of (neutrino) Yukawa  couplings $Y^{BA}_{\nu}$.  
Both terms are $SU(2)_L\times U(1)_Y$ invariant and  even renormalizable. 
We can consider then the diagram shown in Fig. \ref{fig:dim5op_gen}.
At the electroweak scale $v$, if $M_{Maj}\gg v$, we obtain the effective 
interaction shown in Fig. \ref{fig:dim5op}
described by the operator
\begin{equation}
\epsilon_{ij}H_i~l^A_j~Y^{DA}_{\nu}(M^{-1}_{Maj})^{DC}Y^{CB}~\epsilon_{ij}
H_il_j^B\label{32}
\end{equation}
We recognize the previously introduced operator (\ref{16}) with
\begin{equation}
\frac{\lambda_{AB}}{M}=(Y^T_{\nu}M^{-1}_{MAJ}Y_{\nu})_{AB}
\label{33}
\end{equation}

The see-saw mechanism with a new mass scale $M$ is the most compelling 
explanation of the smallness of the neutrino masses \cite{CHPO,ALFE}. Indeed
\begin{description}
\item[i)] the smallness of $m_\nu$ is then related to its zero electric charge
\item[ii)] the smallness of $m_\nu$ is also related to lepton number 
violation at the scale M
\item [iii)] with $m_\nu\sim Y^2\frac{v^2}{M}$, $v=240$~GeV and for 
$Y\sim{\cal O}(1)$ we get $m_\nu \sim (0.01\div0.1)$~eV for 
$M\sim(10^{15}\div10^{13})$~GeV
So, the scale M is close to the GUT scale.
\item[iv)] $\nu^c$ completes the spinor representation of S0(10)
\item[v)] heavy $\nu^c$ can play important role in baryogenesis 
via leptogenesis.
\end{description}
\vskip0.3cm

{\bf 3.4 Hierarchy problem in the SM: hint for a new {\it low} mass scale?}
\vskip0.2cm

Quantum corrections to the Higgs potential mass parameter $m^2$ in 
eq.~(\ref{8}) in the SM are quadratically divergent. If the SM is an
effective low energy theory and has a cut-off at some mass scale $M$
of new physics, $\Lambda_{SM}\sim M$, it means then that quantum corrections
to $m^2$ are quadratically dependent on the new mass scale present in the
underlying more fundamental theory. When $M\gg M_Z$ this is very unnatural
even if $m^2$, that is $M_Z$, remains a free parameter of this 
underlying theory, and particularly difficult to accept if in the 
underlying theory $m^2$ is supposed to be fixed  or indeed is fixed by some
more fundamental considerations (as e.g. in supersymmetric and 
Little Higgs models, respectively). The latter is necessary if the
underlying theory is to predict the scale of the electroweak symmetry 
breaking in terms of ``more fundamental'' parameters and, generically,
in terms of its own cut-off $\Lambda_{\rm New}$. Thus, for naturalness of
the Higgs mechanism in the SM there should exist a new mass scale 
$M\simgt M_Z$, say only order of magnitude higher than $M_Z$ and better 
understanding the mechanism of the electroweak symmetry breaking is, 
hopefully, a bridge to new physics that will be explored at the LHC.

For a more quantitative discussion of this so-called hierarchy problem
we recall that in general in a field theory with a cut-off $\Lambda$ 
and some scalar field(s) $\phi$ that can acquire VEV(s) the 1-loop 
effective potential is
\begin{eqnarray}
\Delta V_{\rm 1-loop}(\phi)={1\over2}\int^\Lambda{d^4k\over(2\pi)^4}~{\rm STr}
\ln\left[k^2-{\cal M}^2(\phi)\right]
 = c\Lambda^4 + c^\prime\Lambda^4\ln\Lambda^2\nonumber\\
+{1\over32\pi^2}\Lambda^2{\rm STr}{\cal M}^2(\phi)
+{1\over64\pi^2}{\rm STr}{\cal M}^4(\phi)\ln{{\cal M}^2(\phi)\over\Lambda^2}
+\dots
\label{eqn:Voneloop}
\end{eqnarray}
where Str${\cal M}^2(\phi)=$Tr$[(-)^FM^2(\phi)]$ with $F$ - the fermion number
operator and ${\cal M}^2(\phi)$ is the full $\phi$-dependent mass matrix for 
all fields of the theory. The first terms in the expansion are  the 
$\phi$-independent contribution to the vacuum energy. We are interested 
in quantum corrections $\delta m^2$ to the mass parameter $m^2$ of the 
$\phi$ field potential. They are obtained by expanding 
\begin{eqnarray}
{\rm STr}{\cal M}^2(\phi)=c_2\phi^2+\dots~,\phantom{aaa}
{\rm STr}{\cal M}^4(\phi)=c_4\phi^2+\dots~,\label{eqn:STRsexps}
\end{eqnarray}
The corrections proportional to $c_2$ (to $c_4$) are in general quadratically
(logarithmically) dependent on the cut-of scale $\Lambda$.
In the SM with a cut-of $\Lambda_{\rm SM}$ we find 
\begin{eqnarray}
\delta m^2=\left(
{\partial^2\Delta V_{\rm 1-loop}(\phi)\over\partial\phi^2}\right)_{\rm min}
={3\over64\pi^2}\left(3g_2^2+g_1^2+\lambda -8y_t^2\right)
\Lambda^2_{\rm SM}+\dots~,
\end{eqnarray}
If the SM was the correct theory up to the mass scale suggested by the
see-saw mechanism, $\Lambda_{\rm SM}\sim M_{GUT}$ 
\begin{eqnarray}
|\delta m^2|\sim10^{30}~{\rm GeV}^2\sim10^{26}M_W^2{\rm ~!}\nonumber
\end{eqnarray}
Clearly, this excludes the possibility of understanding the magnitude of 
Fermi constant $G_F\sim M_W^{-2}$ in any sensible way. We also see that for
$|\delta m^2|\sim M_W^2$, $10M_W^2$, $100M_W^2$ one needs 
$\Lambda_{\rm SM}\simlt0.5$~TeV, $\simlt1$~TeV, $\simlt6$~TeV, respectively.
If the scale of new physics is in the above range it should be discovered
at the LHC.

However, for a solution of the hierarchy problem it is not enough to 
have a low physical cut-off scale of the SM. The deeper theory has its
own cut-off scale $\Lambda_{\rm New}$ and the dependence on it of
$\delta m^2$ calculated in this deeper theory should be mild enough, in
order not to reintroduce the hierarchy problem for the electroweak 
scale.\footnote{The hierarchy of some other (new) scales is nevertheless
usually present.}

Many theoretical ideas have been proposed for solving the hierarchy 
problem of the electroweak scale. In supersymmetric extensions of the SM
the dependence on their own cut-off scales $\Lambda_{\rm New}$ is
only logarithmic because the quadratic divergences cancel out at any
order of the perturbation expansion. Since the effective potential 
$V_{\rm 1-loop}(\phi)$ depends only on $\ln\Lambda_{\rm New}$ (to any order 
of the perturbation expansion) the scale $\Lambda_{\rm New}$ can be as high
as the Planck scale. The quadratic dependence on the SM cut-off scale
$\Lambda_{\rm SM}$, that is on the mass scale $M_{\rm SUSY}$ of the
superparticles, shows up in 
\begin{eqnarray}
{\rm STr}{\cal M}^4(\phi)=f(M^2_{\rm SUSY})~\phi^2+\dots
\end{eqnarray}
and more explicitly, at the 1-loop as\footnote{The formula 
(\ref{eqn:dmSUSY}) applies in fact to $m^2_{H_2}$ which for 
$\tan\beta\simgt5$ is the most important for electroweak symmetry breaking.}
\begin{eqnarray}
\delta m^2={1\over16\pi^2}\left(3g^2_2+g^2_1-12 y^2_t\right)M^2_{\rm SUSY}
\ln{\Lambda^2_{\rm New}\over M^2_{\rm SUSY}}~,\label{eqn:dmSUSY}
\end{eqnarray}
where we have replaced all soft supersymmetry breaking  mass terms
including the Higgs boson mass by $M_{\rm SUSY}$.

Eq.~(\ref{eqn:dmSUSY}) shows that in supersymmetric models the electroweak 
scale is calculable in terms of the known coupling constants and the 
(unknown) scales $M_{\rm SUSY}$ and $\Lambda_{\rm New}$. For a natural 
solution to the hierarchy problem of the electroweak scale $M_{\rm SUSY}$
has to be low, say $M_{\rm SUSY}\simlt{\cal O}(10)M_W$. However, a new
very difficult question appears about the hierarchy 
$\Lambda_{\rm New}/M_{\rm SUSY}$. This is the question about the mechanism 
of supersymmetry breaking. In gravity mediation scenarios 
$\Lambda_{\rm New}\sim M_{\rm Pl}$. In gauge mediation scenarios 
$\Lambda_{\rm New}$ is low but it is a new, introduced by hand, scale. 

Other ideas for solving the hierarchy problem of the electroweak scale 
are more ``pragmatic''. Focusing on the scenarios with some predictive power,
their general structure is the following: the low energy electroweak theory 
(but not necessarily the renormalizable SM) is embedded in a bigger one
with a characteristic mass scale $\Lambda_{\rm SM}\sim M\sim{\cal O}(1$~TeV).
The new physics is under perturbative control up to its cut-off
$\Lambda_{\rm New}\simgt{\cal O}(10$~TeV), high enough to avoid any
conflict with precision electroweak data (to be discussed later). For 
such scenarios with $\Lambda_{\rm New}\simgt{\cal O}(10$~TeV) to be useful for
solving the electroweak hierarchy problem the dependence of $\delta m^2$
on $\Lambda_{\rm New}$ calculated in the extended theory has to be weak enough.
This is obtained by ensuring that at least 1-loop contribution to the 
effective potential (\ref{eqn:Voneloop}) have no quadratic dependence on 
$\Lambda_{\rm New}$:
\begin{eqnarray}
(\delta m^2)_{\rm 1-loop}=0\cdot\Lambda^2_{\rm New}
+{\cal O}(\ln\Lambda_{\rm New})+{\rm const}.
\end{eqnarray}
E.g. in the Little Higgs models the vanishing of the $c_2$ in 
${\rm STr}{\cal M}^2(\phi)$ in eq.~(\ref{eqn:STRsexps}) is ensured by 
cancellation between contributions from particles of the same statistics. 
Such models predict the existence of new quark-like fermions and gauge bosons
with masses $\sim M$. In these models, the quadratic dependence of 
$\delta m^2$ on $\Lambda_{\rm New}$ is 
present in higher order of the perturbation expansion  but it is suppressed
by loop factors. The tree level Higgs mass parameter $m^2$ usually vanishes
$m^2(M)_{\rm tree}=0$ as e.g. the Higgs boson is a Goldstone
boson of some bigger (approximate) symmetry spontaneously broken at the 
scale $M$, i.e. $M$ is identified with the ``decay constant'' and 
$\Lambda_{\rm New}\sim4\pi M$. 
The electroweak symmetry is broken by quantum
corrections. The electroweak scale is then predicted e.g. 
$M_W=M_W($couplings,$M,\Lambda_{\rm New})$ with a mild dependence on
$\Lambda_{\rm New}\simgt{\cal O}(10$~TeV) at which new unknown strong 
interactions set on. The crucial role is played by the new physics 
parameter $M$. In judging the plausibility of such ideas it is
worth remembering our remarks in 3.1.
\vskip0.3cm

{\bf 3.5 New low mass scale and precision electroweak data}
\vskip0.2cm

The presence of new physics at low energy scale, $M\sim1$~TeV raises
the question on its contribution to the electroweak observables. We can
address this question in a model independent way if again we assume
the Appelquist-Carazzone decoupling scenario, i.e. renormalizable SM and
corrections to it from new physics as higher dimension operators:
\begin{eqnarray}
{\cal L}_{\rm eff}={\cal L}_{\rm SM}+\sum_{\hat O^{n+4}_i}{c_i\over M^n}
\hat O^{n+4}_i~.\label{eqn:Leff}
\end{eqnarray}
This time we are interested in operators which contribute to the electroweak
observables. Such operators are necessarily of dimension $n\geq6$. One can 
classify various contributions from new physics according to the value of the 
coefficients $c_i$ in the Lagrangian (\ref{eqn:Leff}): $c_i\sim{\cal O}(1)$ 
for new tree-level contributions or contributions from new strong interacting 
sector; $c_i\sim{\cal O}(1/16\pi^2)$ for contributions from perturbative new 
physics at 1-loop. Fitting the electroweak observables one obtains limits on 
${c_i\over M^2}$. Strictly speaking, the limits are applicable to new physics 
which gives renormalizable SM as its low energy limit but the results are 
also indicative for new mass scales in models like e.g. Little Higgs, in 
which the Appelquist-Carazzone decoupling does not work. In any case, the 
constraints on such models from the electroweak observables can be discussed 
model by model.

The task of using electroweak data to put limits on the new scale $M$ is
greatly facilitated by the expectation that the dominant corrections from
new physics will show up as corrections to the gauge boson self-energies
(the so-called oblique corrections). This expectation is based on the 
experience gained in the SM and with its various hypothetical extensions. 
Several authors have obtained limits on the scale $M$ under plausible 
assumption that the main corrections to the SM fits appear in gauge boson 
self-energies

A comment is in order here. Gauge boson self-energies are not gauge 
invariant objects. The vertex and box contributions contain also pieces 
which are independent of the external legs. They cancel the gauge dependence 
of the full gauge boson self-energies and restore Ward-identity. Therefore, 
strictly speaking, there is an ambiguity in extracting the gauge boson self 
energies from fits to experimental data. Fortunately, the vertex and box
contributions are in the SM in the commonly used gauges much smaller than the
gauge boson self energies and the neglection of gauge dependence of the latters
in obtaining experimental information about their magnitude seems to be a 
reasonable approach.

In the SM, with unbroken electromagnetic $U(1)$ gauge symmetry there are four
independent gauge boson self-energies $\Pi_{ij}(q^2)$:
\begin{eqnarray}
\int d^4x~e^{-iq\cdot x}\langle J^\mu_i(x)J^\nu_j(0)\rangle=
ig^{\mu\nu}\Pi_{ij}(q^2)+q^\mu q^\nu~{\rm term}
\end{eqnarray}
For instance, we can 
take $\Pi_{\gamma\gamma}$, $\Pi_{3\gamma}$, $\Pi_{33}$ and $\Pi_{11}$ 
as independent quantities ($i=1,3$ are $SU(2)_L$ indices; the QED Ward
identity implies $\Pi_{11}(q^2)=\Pi_{22}(q^2)$. In the limit of $q^2\ll M^2$
we can expand
\begin{eqnarray}
&&\Pi_{\gamma\gamma}(q^2)\approx q^2\Pi^\prime_{\gamma\gamma}(0)\nonumber\\
&&\Pi_{3\gamma}(q^2)\approx q^2\Pi^\prime_{3\gamma}(0)\nonumber\\
&&\Pi_{33}(q^2)\approx\Pi_{33}(0)+q^2\Pi^\prime_{33}(0)\\
&&\Pi_{11}(q^2)\approx\Pi_{11}(0)+q^2\Pi^\prime_{11}(0)\nonumber
\end{eqnarray}
($\Pi_{\gamma\gamma}(0)=0$ by QED Ward identity; the only non-zero 
contribution to $\Pi_{3\gamma}(0)$ comes from the $W^\pm$-charged
Goldston boson loop).
Thus, oblique corrections to the electroweak observables are to a good
approximation parametrized by six constants. Three of them (or three linear
combinations) are fixed in terms of $\alpha$, $M_Z$ and $G_F$ by the 
renormalization procedure. In the remaining three combinations the UV 
divergences must cancel.\footnote{The finitness of the gauge sector 
contribution to $S$, $T$ and $U$ requires the inclusion of the terms
with $\Pi_{3\gamma}(0)$.}
One usually defines \cite{ALBA,PETA}
\begin{eqnarray}
&&\alpha T\equiv{e^2\over s^2c^2M_Z^2}\left[\Pi_{11}(0)-\Pi_{33}(0)\right]
={\Pi_{WW}(0)\over M_W^2}-{\Pi_{ZZ}(0)\over M_Z^2}\nonumber\\
&&\alpha S\equiv4e^2\left[\Pi^\prime_{33}(0)-\Pi^\prime_{3\gamma}(0)\right]
\propto\Pi^\prime_{3Y}(0)\\
&&\alpha U\equiv4e^2\left[\Pi^\prime_{11}(0)-\Pi^\prime_{33}(0)\right]\nonumber
\end{eqnarray}
It is clear from their definition that the parameters $S$, $T$ and $U$ have
important symmetry properties: $T$ and $U$ vanish in the limit of unbroken
custodial $SU(2)_V$ symmetry. The parameter $S$ vanishes when $SU(2)_L$
is unbroken; unbroken $SU(2)_V$ is not sufficient for vanishing of $S$ 
because $S\propto\Pi^\prime_{3Y}(0)=\Pi^\prime_{3L,3R}(0)+\Pi^\prime_{3L,B-L}$
(the decomposition is labelled by the $SU(2)_L\times SU(2)_R\times U(1)_{B-L}$
quantum numbers) and $\mathbf{3}_L\times\mathbf{3}_R=\mathbf{1}+\mathbf{5}$
under $SU(2)_V$. 

It turns out that in the SM the quantum corrections to the $\rho$ parameter 
defined in section 2 as a ratio of physical (and measured) observables are 
to a very good approximation given by $\alpha T$:
\begin{eqnarray}
\Delta\rho\equiv\rho-1={M_W\over M_Z\cos\theta}-1\approx\alpha T
\label{eqn:drho}
\end{eqnarray}
and, according to the eq.~(\ref{eqn:drhoSM}), depend quadratically on the
top quark mass and logarithmically on the SM Higgs boson mass.
Eq.~(\ref{eqn:drho}) is a good approximation because other corrections to
$\rho$ (vertex corrections) are in the SM negligibly small. Similarly, in 
the SM
\begin{eqnarray}
\alpha S={e^2\over48\pi^2}\left(-2\ln{m^2_t\over M^2_Z}+\ln{M_h^2\over M^2_Z}
\right)~.\label{eqn:SSM}
\end{eqnarray}

It was discussed in section  2.3 that the SM quantum corrections agree
excellently with electroweak data. The only free parameter in the
fits of the SM to these data is the Higgs boson mass. The main part of 
this dependence  enters through the $\rho$ parameter and the data favour 
negligible contribution to $\rho$ from $\ln(M_h/M_W)$
(see eq.~(\ref{eqn:drhoSM})). The fits 
give $M_h\approx{\cal O}(100$~GeV), with a big error since the dependence 
of the fits on the Higgs boson mass is only logarithmic. Such fits determine 
the values of parameters $S_{SM}$, $T_{SM}$ and $U_{SM}$ for the best fitted 
value of $M_h$.

We can discuss the room for new physics contribution to the electroweak
fits by writing more generally:
\begin{eqnarray}
&&T=T_{SM}(M_h)+\Delta T~,\nonumber\\
&&S=S_{SM}(M_h)+\Delta S~,\\
&&U=U_{SM}(M_h)+\Delta U~,\nonumber
\end{eqnarray}
where $T_{SM}(M_h)$ etc. is the SM contribution for some fixed value
of the Higgs boson mass.

A fit to the data gives now some values for $\Delta T$, $\Delta S$ and
$\Delta U$ as a function of the assumed value of $M_h$ and shows
how much room we have for new physics for different values of $M_h$.
It is clear that for, say, $M_h=115$~GeV such fits give $\Delta T$, 
$\Delta S$ and $\Delta U$ consistent with zero and the only room for new
physics is in the errors of the fitted values of these quantities.
For larger values of $M_h$ we have more room for new physics contributions.
As we can see from eqs.~(\ref{eqn:drhoSM}) and (\ref{eqn:SSM}) it must be
positive to $T$ and negative to $S$ to balance the contribution of the 
larger Higgs boson mass. The fitted values of $\Delta T$, $\Delta S$ and
$\Delta U$ for different values of $M_h$  give limits on the coefficients
$c_i/M$ of the dimension six operators that contribute to $\Delta T$, 
$\Delta S$ and $\Delta U$. 

Several interesting conclusions have been reached in such studies
\cite{HAKO}. First of 
all, independently of the assumed value of $M_h$, for $c_i\sim{\cal O}(1)$
the fits to the electroweak data give a lower limit $M\simgt{\cal O}(4$~TeV)
This limit is reached only for very correlated signs of the coefficients $c_i$.
Thus, qualitatively speaking, any new physics with $M\simlt{\cal O}(10$~TeV)
must be perturbative and cannot contribute at the tree level to be consistent
with electroweak data. This strongly suggests a perturbative solution to the
hierarchy problem.

Secondly, if new physics is indeed perturbative and shows up only at loop
level, then the fits give $M_h\simlt240$~GeV independently of the value
of the scale $M$. 

Finally, if $M_h\simgt300$~GeV then new physics with the scale 
$M\sim10\div30$~TeV and with $c_i\sim{\cal O}(1)$ (i.e. strongly interacting)
is actually needed! Thus, experimental discovery of the Higgs boson and
the determination of its mass will be a strong hint about the kind of new
physics one may expect.

\section{Supersymmetric extensions of the Standard Model}

In the rest of these lectures we discuss supersymmetric models as at present 
the most complete theoretical framework going beyond the Standard Model
\cite{KANE1}. It 
has quite a few attractive features and also a number of difficulties, so 
that the full success of the SM is not automaticaaly recovered. We
have already mentioned some of them in several places in these lectures but 
here we collect them together and extend our discussion. We shall not
discuss one fundamental unresolved issue for supersymmetry which is the
mechanism of spontaneous supersymmetry breaking and its transmission to the
the SM sector. On the phenomenological side, the related problem is that of 
new and potentially dangerous sources in the soft supersymmetry breaking 
parameters of the FCNC and CP violating transitions. For a review of
all these aspects of supersymmetric and supergravity models see e.g. 
\cite{SUGRA,GAGAMASI,KANEetAL}.
\vskip0.3cm

{\bf 4.1 Precision electroweak data}
\vskip0.2cm

As discussed in section 3.4
supersymmetric models like the Minimal Supersymmetric Standard Model (MSSM) 
or its simple extensions satisfy a very important criterion of calculability. 
Most of the structure of the Standard Model is built into them, so the 
renormalizable Standard Model is their low energy approximation. 
Supersymmetric models are easily consistent \cite{paris} with the electroweak 
data since the supersymmetric quantum corrections to the Standard Model fits
are suppressed by powers of the mass scale $M_{SUSY}$ of supersymmetric 
particles and for $M_{SUSY}>{\cal O}(500)$~GeV are well below experimental 
errors (in particular, the custodial symmetry breaking by the sfermion masses
is sufficiently suppressed). Thus, the predictive power of the Standard Model 
remains intact and its success is not accidental.
\vskip0.3cm

{\bf 4.2 The electroweak symmetry breaking}
\vskip0.2cm

Supersymmetric models solve the hierarchy problem of the electroweak scale. 
In the limit of unbroken supersymmetry the quadratically divergent quantum 
corrections to the Higgs mass parameter are absent in any order of 
perturbation theory. When supersymmetry is softly broken  by a mass scale $M$, 
the superpartners get their masses from the electroweak breaking and from the
supersymmetry breaking mass terms $\sim M$. They decouple at energies smaller
than $M$ and the quadratically divergent Standard Model contribution to the 
Higgs mass parameter is cut-off by $M$ and, therefore, depends quadratically 
on M. Thus, the hierarchy problem of the electroweak scale disappears
if $M\simlt{\cal O}(1)$~TeV. The cut-off to a supersymmetric theory can be 
as high as the Planck scale and ``small'' scale of the electroweak symmetry 
breaking is still natural.

The electroweak symmetry breaking may be triggered by radiative corrections 
to the Higgs potential:
\begin{equation}
(\delta m^2_{H_2})_{\rm 1-loop}\sim -{\cal O}(0.1)~M^2\ln{\Lambda\over M}
\label{4.1}
\end{equation}
This formula follows from eq.~(\ref{eqn:dmSUSY}). If we assume that 
$m^2_0\sim M^2$, i.e. that the tree level Higgs mass parameter is 
approximately equal to the soft supersymmetry breaking scale the radiative 
electroweak symmetry breaking ($(m^2_{H_2})_{\rm1-loop}<0$) is triggered 
by the large top quark Yukawa coupling, hidden in the numerical factors 
of eq.~(\ref{4.1}). With the Higgs boson self-interactions fixed by the 
gauge couplings of the Standard Model
\begin{equation}
\lambda\phi^4\rightarrow g^2\phi^4
\label{4.2}
\end{equation}
one obtains the correct prediction for the electroweak scale for 
$\Lambda\sim M_{GUT,~Pl}$. This nicely fits
with unification of the gauge couplings.
\vskip0.3cm

{\bf 4.3 The mass of the lightest Higgs boson}
\vskip0.2cm

Supersymmetric models typically restrict the couplings in the Higgs potential 
and give strong upper bounds on the mass of the lightest Higgs particle
\cite{KANE2}. In the minimal model the Higgs boson self-coupling comes from 
the D-terms and its self-coupling  is the gauge coupling, eq.(\ref{4.2}). 
Therefore, at the tree level
\begin{equation}
M_{\rm Higgs}<M_Z\approx91 ~{\rm GeV}
\label{4.3}
\end{equation}
There are large quantum corrections to this result. They depend quadratically 
on the top quark mass and logarithmically on the stop mass scale 
$M_{\tilde t}\sim M_{\rm SUSY}$:
\begin{equation}
M^2_{\rm Higgs}=\lambda v^2
\end{equation}
where $\lambda$ is given by
\begin{equation}
\lambda={1\over8}(g_2^2+g_1^2)\cos^22\beta+\Delta\lambda~,\phantom{aaaa}
{\rm with} \phantom{aa}
\Delta\lambda={3g^2_2\over8\pi^2}{m_t^4\over v^2M^2_W}
\ln{M^2_{\tilde t}\over m^2_t}~.\label{eqn:MSSMHigmass}
\end{equation}
The present experimental limit $M_{\rm Higgs}>114$ GeV requires 
$M_{\tilde t}\simgt500$~GeV and for $M_{\tilde t}<1$~TeV, 
$M_{\rm Higgs}<130$~GeV. The closer the Higgs mass would be to 
the present experimental limit, the better it would be for the ``naturalness'' 
of the electroweak scale. Clearly, in the MSSM, the tunning in the Higgs 
potential depends exponentially on the Higgs mass and one may eventually have 
some tension here (see eqs.~(\ref{eqn:MSSMHigmass}), (\ref{eqn:dmSUSY})).

One can depart from the minimal model and relax the bound on the Higgs mass. 
For instance, with an additional chiral superfield which is a Standard Model 
singlet, one may couple the singlet to the Higgs doublets and get
additional contributions to the Higgs self-coupling. Explicit calculations 
show that in such and other models, with $M\simlt1$~TeV, 
the bound on the Higgs mass cannot be raised above $\sim 150$~GeV if one 
wants to preserve perturbative gauge coupling unification.
\vskip0.3cm

{\bf 4.4 Gauge coupling unification}
\vskip0.2cm

It is well known \cite{RABY} that in the framework of the MSSM with 
degenerate sparticle spectrum characterized by $M_{SUSY}\approx1$~TeV the 
three experimentally measured gauge couplings unify with high precision at 
the scale $M_{GUT}\sim10^{16}$~GeV. This gives support to perturbative new 
physics at ${\cal O}(1$~TeV). Supersymmetry and the idea of grand unification 
(see section 3.2) mutually strengthen their attractiveness.

A closer look at the unification is interesting.
One may ask how precise is the unification when the superpartner masses 
are not degenerate and different from 1~TeV. It has been understood that
even for nondegenerate superpartner spectrum the superpartner mass dependence 
of the RG evolution of the gauge couplings can be described to a very good 
approximation by a single effective parameter $T$. The superparticle 
threshold effects are correctly included in the supersymmetric 1-loop RGE
whose running starts at $T$, with the SM RG equations used
below the scale $T$. For consistency, 2-loop running should also be included. 

$T$ depends strongly on the higgsino ($\mu$) and gaugino ($M_i$) mass 
parameters and  much weaker on the sfermion masses. Exact unification of the 
measured gauge couplings requires $T\approx1$~TeV, i.e. the higgsino and the 
gaugino physical masses $\sim1$~TeV if degenerate. However, a more plausible 
assumption that the parameters $\mu$ and $M_i$ are approximately degenerate 
and  $\sim1$~TeV at the GUT scale gives $T\approx100$~GeV because of strong 
renormalization effects. Thus, a realistic spectrum does not give exact 
unification and one may wonder about the accuracy of unification in the MSSM. 

In order to define what we understand by 'successful unification' let us 
first recall the one-loop renormalization group equations in the SM and MSSM. 
At one-loop the gauge couplings $\tilde{\alpha}_i$ of the three group 
factors of $G_{\rm SM}$ run according to the equations: 
\begin{eqnarray} 
\label{eqn:runningsm}
{1\over\tilde{\alpha}_i(Q)} = 
{1\over\tilde{\alpha}_i(M_Z)} - {b_0^{(i)}\over 2\pi} 
\ln\left({Q\over M_Z}\right)  + \delta_i  
\end{eqnarray}
Here, $1/\tilde{\alpha}_i(M_Z)=(58.98\pm0.04, ~29.57\pm0.03, ~
8.40\pm0.14)$ are the experimental values of the gauge couplings 
at the $Z^0$-pole and $b_0^{(i)}$ are the one-loop coefficients of the 
relevant beta-functions. They read 
$b_0=({1\over10}+{4\over3}N_g, -{43\over6}+{4\over3}N_g, -11+{4\over3}N_g)$
in the SM and $b_0=({3\over5}+2N_g, -5+2N_g, -9+2N_g)$ in the MSSM, where 
$N_g$ is the number of generations. Threshold corrections (e.g. from heavy 
GUT gauge bosons) are represented by the parameters $\delta_i$. As explained
earlier using the MSSM RG equations directly from the electroweak scale
for $T\approx M_Z$ means that the supersymmetric threshold corrections 
corresponding to a realistic mass spectrum are properly included.

In the bottom-up approach one can speak about the gauge coupling unification 
if in some range of scales $Q$ the couplings defined 
by eq.~(\ref{eqn:runningsm}) with, in general, $Q$-dependent $\delta_i(Q)$ 
can take a common value $\alpha_i(Q)=\alpha_{\rm GUT}$ for reasonably small 
values of $\delta_i(Q)$ (compared to $\alpha^{-1}_{\rm GUT}$).\footnote{
Whether there exists a unified model able to provide such values of 
$\delta_i(Q)$'s is a different question.}  
The condition for the unification can be succintly written as
\begin{eqnarray} 
\epsilon_{ijk}\left({1\over\tilde{\alpha}_i(M_Z)} + \delta_i\right)
(b_0^{(j)}- b_0^{(k)}) = 0 
\end{eqnarray}
Putting in the experimental values for $\alpha_i(M_Z)$ and the 
beta-function coefficients we get:  
\begin{eqnarray} 
 -41.1 + 3.8 \delta_1 -11.1 \delta_2 + 7.3 \delta_3 =0 & & ({\rm SM}) 
\nonumber\\
-0.9 +4 \delta_1 - 9.6 \delta_2 + 5.6 \delta_3 =0 & &({\rm MSSM})  
\end{eqnarray}

We see, that to achieve the gauge coupling unification at the one-loop 
level we need the threshold corrections $\delta_i$ to be of order 10\% 
$\alpha_{\rm GUT}^{-1}$ in the SM, while in the MSSM we need  only 
$\delta_i\sim$1\% $\alpha_{\rm GUT}^{-1}$. In the MSSM once the two loop 
effects are inluded one needs $\delta_i$'s by factor 2 larger. The 
unification of the gauge cuplings in the MSSM is indeed very precise: it 
admits (and requires) only 2\% threshold corrections from the GUT physics. 
These 2\% corrections give 10\% effect on $\alpha_s$ at $M_Z$ scale, but
the precision of unification in the MSSM should be judged by the necessary 
for exact unification threshold corrections at the GUT scale. 

Unification of the gauge couplings does not necessarily imply the 
standard GUT theories with all their problems, like spontaneous breaking
of the GUT gauge group by VEVs of some Higgs fields, the doublet-triplet
splitting problem, etc. Many different solutions have been proposed.

With the threshold corrections of the right order of magnitude, 
the  unification scale can be estimated from the equation:  
\begin{eqnarray} 
\label{u1} 
{1\over\alpha_1(M_Z)}-{1\over\alpha_2(M_Z)}
-{1\over 2\pi}(b_0^{(1)}- b_0^{(2)})\ln\left({M_{\rm GUT}\over M_Z}\right) 
+ (\delta_1 - \delta_2) =0 
\end{eqnarray} 
For the sake of concreteness, we assume here that 
$\delta_1=\delta_2=0$ and that all 
threshold corrections are accounted for by $\delta_3$ (thus, the unification 
point is assumed to be where  $\alpha_1$ and $\alpha_2$ intersect).  
Putting in the experimental numbers and the beta-function coefficients we get 
$M_{\rm GUT}\approx1\times10^{13}$ GeV
in the SM and $M_{\rm GUT}\approx 2\times10^{16}$ GeV in the MSSM.

The scale of unification in the MSSM is determined very precisely to be
in the range $(2\div4)\times10^{16}$~GeV. This is interesting because it is 
very close to the reduced Planck scale $M_{\rm Pl}=2\times10^{18}$~GeV
and could be considered as evidence for unification including 
gravity.\footnote{In string theories without the stage of Grand Unification 
below the compactification scale $M_S$ the couplings unify at $M_S$ which
e.g. in weakly coupled heterotic string is about factor 5 below $M_{\rm Pl}$.}
But one to two orders of magnitude difference between the two scales needs 
some explanation.  Of course, new particles in incomplete $SU(5)$ 
representations would alter the running and could push the unification scale 
closer to the Planck scale. However, one must not destroy the precision of 
the unification by new threshold corrections,
so this possibility looks very fine-tuned and {\sl ad hoc}. An 
interesting possibility would be to unify the three gauge interactions 
with gravity at $M_{GUT}\sim10^{16}$~GeV by changing the energy dependence
of the gravity coupling. This is possible if gravitational interactions
(and only they) live in more than 3 spatial dimentions. The effective
four-dimensional Planck constant is then 
\begin{eqnarray} 
M^2_{\rm Pl}=M^{2+n}_{{\rm Pl}~(4+n)}R^n~,
\end{eqnarray}
where $n$ is the number of extra dimensions, $R$ is their compactification
radius and $M^{2+n}_{{\rm Pl~(4+n)}}$ is the Planck scale in $4+n$ dimensions
which we would like to take equal to $M_{GUT}\sim10^{16}$~GeV.
For $n=1$ (like in the $M$-theory of Horava and Witten) we get
$1/R\sim10^{14}$~GeV.

\vskip0.3cm

~~~~~{\bf 4.5 Proton decay}
\vskip0.2cm

In the SM the baryon number is (perturbatively) conserved since there are no
renormalizable couplings violating this symmetry (see section. 2.5). 
Experimental search for proton decay, e.g $p\rightarrow e^+\pi^0$,
$p\rightarrow K^+\nu$ is one of the most fundamental tasks for particle 
physics. The present limit is $\tau_p>10^{33}$ years. If the SM is only
an effective low energy theory the remnants of new physics should show up 
as non-renormalizable corrections to the SM. The lowest dimension operators
for the proton decay (with the particle spectrum of the SM) is the set of
dimension 6 operators of the form
\begin{eqnarray}
\hat O^{(6)}_i={c^{(6)}_i\over M^2_{(6)}} qqql\label{eqn:dim6pdec}
\end{eqnarray} 
For such operators for the proton lifetime we get
\begin{eqnarray}
{1\over\tau_p}={[c^{(6)}_i]^2\over16\pi}{m_p^5\over M^4_{(6)}}
\nonumber
\end{eqnarray} 
The limit $\tau_p>10^{33}$ years gives then 
$M_{(6)}>\sqrt{c^{(6)}_i}\times10^{16}$~GeV (this is only a very rough
estimate which neglects strong interaction effects). Any new physics with 
lower mass scale that could lead to proton decay should be coupled with 
$c^{(6)}_i\ll1$. For instance, for $c^{(6)}_i\sim\alpha_{\rm GUT}\approx1/25$ 
we get $M_{(6)}\simgt\times10^{15}$~GeV which is still too high for the SM 
unification ($M_{\rm GUT}\approx10^{13}$~GeV). 

In supersymmetric extensions of the SM, with softly broken low energy 
supersymmetry there are low mass scalars in the spectrum with masses 
$M\sim{\cal O}(1$~TeV), which may have renormalizable couplings to quarks 
and leptons. 

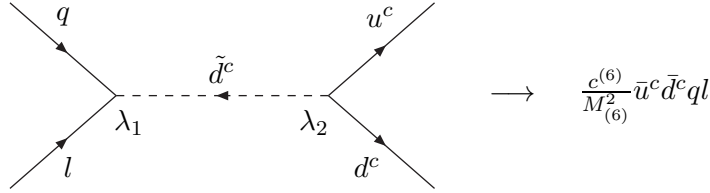
\begin{figure}[htbp]
\begin{center}
\begin{picture}(280,90)(0,0)
\ArrowLine(10,10)(50,45)
\ArrowLine(130,45)(170,10)
\DashArrowLine(130,45)(50,45){3}
\ArrowLine(10,80)(50,45)
\ArrowLine(130,45)(170,80)
\Text(32,18)[]{$l$}
\Text(145,18)[]{$d^c$}
\Text(30,75)[]{$q$}
\Text(150,75)[]{$u^c$}
\Text(55,35)[]{$\lambda_1$}
\Text(125,35)[]{$\lambda_2$}
\Text(90,55)[]{$\tilde d^c$}
\Text(200,45)[]{$\longrightarrow$}
\Text(250,45)[]{${c^{(6)}\over M^2_{(6)}} \bar u^c\bar d^cql$}
\end{picture}
\end{center}
\caption{Diagram generating the dimension 6 operator.}
\label{fig:dim6op_gen}
\end{figure}

Indeed, the most general renormalizable superpotential in the minimal 
supersymmetric model is
\begin{eqnarray}
w=\hat U^c\hat Q\hat H_u + \hat D^c\hat Q\hat H_d + \hat E^c\hat L\hat H_d
+\hat H_d\hat H_u\nonumber\\
+ \hat D^c\hat Q\hat L + \hat E^c\hat L\hat L + \hat U^c\hat D^c\hat D^c
+\hat L\hat H_u~,\label{eqn:gensupot}
\end{eqnarray}
(the coupling constants and the flavour indices are suppressed). The second 
line is also consistent with the  SM gauge symmetry but these interactions
do not conserve baryon and lepton numbers and give renormalizble couplings
of scalars to fermions. After integrating the scalars out one gets dimension 
6 operators as in (\ref{eqn:dim6pdec}) with $M_{(6)}\sim M_{\rm SUSY}$ from 
diagrams shown in Fig.~\ref{fig:dim6op_gen}
and to be consistent with the limit on the proton lifetime we need
$c^{(6)}=\lambda_1\lambda_2<10^{-26}$.

One can forbid the terms in the second line of eq.~(\ref{eqn:gensupot})
by imposing a discrete symmetry, the so-called matter parity
$R_p=(-1)^{3(B-L)}$. Such a symmetry could for instance, be a discrete 
remnant of the gauged $U(1)_{B-L}$ in the $SO(10)$ theory \cite{MARTIN}. 
Matter parity is equivalent to $R$-parity $R=(-1)^{2S+3(B-L)}$ acting
on the component fields, where $S$ is their spin, since Lorentz-invariant
interactions preserve $(-1)^{2S}$. We get then a stable LSP - candidate for 
dark matter in the Universe.

\begin{figure}[htbp]
\begin{center}
\begin{picture}(130,100)(0,0)
\ArrowLine(120,10)(90,20)
\ArrowLine(120,90)(90,80)
\ArrowLine(90,20)(90,50)
\ArrowLine(90,80)(90,50)
\Text(90,50)[]{$\mbox{\boldmath$\times$}$}
\Text(105,50)[]{$\tilde W$}
\DashArrowLine(90,80)(30,50){3}
\DashArrowLine(90,20)(30,50){3}
\Text(60,75)[]{$\tilde l$}
\Text(60,25)[]{$\tilde q$}
\Text(10,75)[]{$q$}
\Text(10,25)[]{$q$}
\Vertex(30,50){2}
\ArrowLine(10,90)(30,50)
\ArrowLine(10,10)(30,50)
\end{picture}
\end{center}
\caption{Loop diagram generating the dimension 6 operator form the 
dimension 5 operator.}
\label{fig:dim5todim6}
\end{figure}
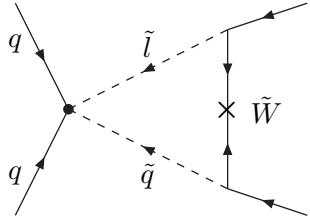

\begin{figure}[htbp]
\begin{center}
\begin{picture}(180,90)(0,0)
\ArrowLine(10,10)(50,45)
\ArrowLine(170,10)(130,45)
\ArrowLine(50,45)(90,45)
\ArrowLine(130,45)(90,45)
\DashArrowLine(10,80)(50,45){3}
\DashArrowLine(170,80)(130,45){3}
\Text(90,45)[]{$\mbox{\boldmath$\times$}$}
\Text(30,18)[]{$q$}
\Text(145,18)[]{$q$}
\Text(60,37)[]{$H_c$}
\Text(118,37)[]{$H_c$}
\Text(30,75)[]{$\tilde q$}
\Text(150,75)[]{$\tilde l$}
\Text(90,55)[]{$M_{H_c}$}
\end{picture}
\end{center}
\caption{Diagram generating the dimension 5 operator for proton decay.}
\label{fig:dim5opfromH}
\end{figure}
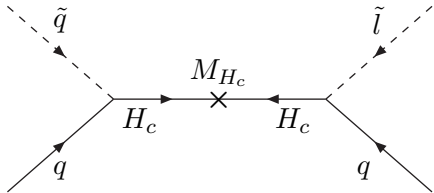

In supersymmetric GUT models, even with $R-$parity imposed, there is still
another source of dangerous contributions to the proton decay amplitudes.
These are the dimension 5 operators 
\begin{eqnarray}
\hat O^{(5)}_i={c^{(5)}_i\over M^2_{(5)}} qq\tilde q \tilde l
\label{eqn:dim5pdec}
\end{eqnarray} 
which when inserted into one loop diagrams with gaugino exchanges
give rise via diagrams shown in Fig.~\ref{fig:dim5todim6} to dimension 6 
operators.
In the effective dimension 6 operator of the form (\ref{eqn:dim6pdec})
one then gets
$c^{(6)}=\alpha_{\rm GUT}c^{(5)}$, $M^2_{(6)}=M_{(5)}M_{\rm SUSY}$. From
$\tau_p>10^{33}$ years and for $M_{\rm SUSY}\sim{\cal O}(1$~TeV), 
$M_{(5)}\sim M_{\rm GUT}\sim10^{16}$~GeV, one gets $c^{(5)}\simlt10^{-7}$.
Thus we need small couplings in the amplitudes generating the dimension 5
operators. In SUSY GUT's dimension 5 operators originate from the exchange 
of the colour triplet scalars present in the Higgs boson GUT multiplets as 
shown in Fig.~\ref{fig:dim5opfromH},
so $c^{(5)}\sim Y^2$ (Yukawa couplings for the quarks of the first two 
generations) and $M_{(5)}\sim M_{H_c}\sim M_{\rm GUT}$. 

Given various uncertainties, e.g. in the unknown squark, gaugino and heavy
Higgs boson mass spectrum, such contributions in supersymmetric GUT models 
are marginally consistent with the experimental limits on the proton 
lifetime, particularly in models more complicated than the minimal 
supersymmetric $SU(5)$ model. Concrete classical GUT models are, however, not 
very
attractive and plagued with various problems like e.g. doublet-triplet
splitting problem for the Higgs boson multiplets. There are several
interesting other ideas like unification in (small) extra dimensions
or in string theory \cite{DIN}, which offer the possibility of avoiding those 
difficulties and simultaneously preserving the attractive features of GUT's.
In some of such models proton is stable or its lifetime makes its decays
unobservable experimentally. An interesting question is: what if proton
after all decays but slow enough to rule out classical GUT models? 

\section{Summary}

Thinking about new physics from the perspective of the extremely succesful
Standard Model is very challenging. At present there is no approach that
fully and convincingly incorporates this succes into its structure. 
Focusing on the electroweak symmetry breaking alone, one may wonder if the 
high predictive power of the renormalizable SM for the electroweak observables 
and its perfect agreement with experimental data is significant or partly
accidental. If significant - it supports supersymmetry; if partly accidental 
- we have
more room for various speculations about the mechanism of electroweak 
symmetry breaking. Experiments are needed to put us on the right track,
and hopefully, experiments at the LHC will give us the necessary insight.
\vspace{1cm}

{\bf Acknowledgments:} I thank Emilian Dudas, Gordon Kane, Jan Louis, Hans 
Peter Nilles, Jacek Pawe\l czyk, Fernando Quevedo, Zurab Tavartkiladze and 
especially my close colaborator Piotr Chankowski for many useful conversations 
about physics beyond the Standard Model. My special thanks go to Jihad Mourad 
who patiently reviewed with me the content of these lectures during my visit 
to the Astroparticle and Cosmology Insititute (Paris VII) in June 2004, from 
the perspective of a string theorist. I thank also the Organizers of the 
Cargese School for a very enjoyable time.

\end{document}